\documentclass{aa}  

\usepackage{graphicx}
\usepackage{txfonts}
\begin{document}

   \title{The Intrinsic Distribution of Lyman-$\alpha$ Halos}

   \subtitle{}

   \author{John Pharo
          \inst{1},
          Lutz Wisotzki\inst{1},
          Tanya Urrutia\inst{1},
          Roland Bacon\inst{2},
          Ismael Pessa\inst{1},
          Ramona Augustin\inst{1},
          Ilias Goovaerts\inst{3},
          Daria Kozlova\inst{1},
          Haruka Kusakabe\inst{4,5},
          H\'ector Salas\inst{1},
          Daniil Smirnov\inst{1},
          Tran Thi Thai\inst{3,6}
          \and
          Elo\"{i}se Vitte\inst{5}
          }

   \institute{Leibniz Institut für Astrophysik (AIP),
              An der Sternwarte 16, 14482 Potsdam, Germany\\
              \email{jpharo@aip.de}
              \and
              Univ. Lyon, Univ. Lyon1, ENS de Lyon, CNRS, Centre de Recherche Astrophysique de Lyon UMR5574, 69230 Saint-GenisLaval, France
              \and
              Aix Marseille Universit\'{e}, CNRS, CNES, LAM (Laboratoire d’Astrophysique de Marseille), UMR 7326, 13388 Marseille, France
              \and
              National Astronomical Observatory of Japan (NAOJ), 2-21-1 Osawa, Mitaka, Tokyo, 181-8588, Japan
              \and
              Observatoire de Gen\`{e}ve, Universit\'{e} de Gen\`{e}ve, Chemin Pegasi 51, 1290 Versoix, Switzerland
              \and
              Department of Astrophysics, Vietnam National Space Center, Vietnam Academy of Science and Technology, 18 Hoang Quoc Viet, Hanoi, Vietnam
              }






   \date{}

   \authorrunning{J. Pharo}

 
  \abstract
  {The emission and escape of Lyman-$\alpha$ photons from star-forming galaxies is determined through complex interactions between the emitted photons and a galaxy's interstellar and circumgalactic gas, causing Lyman-$\alpha$ emitters (LAEs) to commonly appear not as point sources but in spatially extended halos with complex spectral profiles. We develop a 3D spatial-spectral model of Lyman-$\alpha$ halos (LAHs) to replicate LAH observations in integral field spectroscopic studies, such as those made with VLT/MUSE. The profile of this model is a function of six key halo properties: the halo- and compact-source exponential scale lengths ($r_{sH}$ and $r_{sC}$), the halo flux fraction ($f_H$), the compact component ellipticity ($q$), the spectral line width ($\sigma$), and the spectral line skewness parameter ($\gamma$). Placing a series of model LAHs into datacubes reflecting observing conditions in the MUSE UDF-Mosaic survey, we test their detection recoverability and determine that $\sigma$, $r_{sH}$, and $f_H$ are expected to have the most significant effect on the detectability of the overall LAH at a given central wavelength and intrinsic line luminosity. We develop a general selection function model spanning a grid of these halo parameters, and with a sample of 145 LAHs with measured halo properties observed in the UDF-Mosaic survey, we derive completeness-corrected, intrinsic distributions of the values of $\sigma$, $r_{sH}$, and $f_H$ for $3<z<5$ LAHs. We present the best-fit functional forms of the distributions, and present as well a $\sigma$ distribution corrected for instrumental line-spread function (LSF) broadening, and thereby show the physical line spread distribution of the intrinsic population. Finally, we discuss possible implications for these distributions to the nature of Ly$\alpha$ emission through the circumgalactic medium, finding that observations may undercount LAHs with extended halo scale lengths compared to the intrinsic population.}


   \maketitle
%

\section{Introduction}

The Lyman alpha emission line (Ly$\alpha$) of hydrogen is one of the most effective and commonly used observational tools in the study of galaxy evolution. High-energy photons from young, massive stars ionize neutral hydrogen in a galaxy's interstellar gas, and when the ionized hydrogen and electron recombine, a Ly$\alpha$ photon is emitted with high probability \citep{partridge1967}. The emitted Ly$\alpha$ photon may be absorbed or scattered by any neutral hydrogen it encounters, and therefore the Ly$\alpha$ photon's travel from its origin in the galaxy's interstellar medium (ISM) through the surrounding circumgalactic medium (CGM) and intergalactic medium (IGM) will be influenced by the properties of the gas it encounters. The path the photon traverses through these media to the point of observation is determined by complicated radiative transfer pathways that depend on density, temperature, composition, and kinematic properties of the several phases of intervening gas \citep{ouchi2020}. Consequently, Ly$\alpha$ emission is a possible tracer of galaxy star formation rate \citep{sobral2019}, characteristics of the baryonic matter in the CGM \citep{muzahid2021, banerjee2023, galbiati2023, lofthouse2023}, and ionization of the IGM \citep{malhotra2006, stark2010, matthee2022, goovaerts2023}, making the detection of Lyman-$\alpha$ emitting galaxies (Lyman-$\alpha$ emitters, or LAEs) a powerful probe of several critical phases of galaxy evolution.

At high redshift, LAEs and their contributions to reionization are analyzed via population studies, such as with the construction of luminosity functions \citep[LFs, ][]{malhotra2004, ouchi2008, finkelstein2012, drake2017, sobral2018, herenz2019, wold2022, thai2023}, which characterize the numerical distribution of LAEs as a function of the Ly$\alpha$ line luminosity in a given redshift epoch \citep[e.g.,][]{johnston2011} and in a representative volume of the universe: $dN = \Phi(L)dLdV$. Through observation of a population of LAEs at given redshifts, this differential LF may then be evaluated to determine a number density distribution of LAEs as a function of luminosity (or other LAE property) for a given redshift epoch. The form this LF takes is commonly parameterized with the Press-Schecter function \citep{press1974, schechter1976}, but may also be estimated non-parametrically.

Analyses that require knowledge of the intrinsic LAE population \textemdash\ luminosity functions, LAE clustering properties \citep{herrero2023a}, cosmic star formation rate density \citep{sobral2019}, etc. \textemdash\ require not just measurement of the observed LAEs, but careful consideration of the completeness of the flux-limited observations \citep[see, for example,][]{drake2017}. The faint LAE population is difficult to consistently detect at high redshift, especially with spectroscopy, and so luminosity-dependent formalizations of the LF will suffer from a detection bias toward intrinsically brighter sources. Many non-parametric methods for estimating the LF have been developed to address this problem. In this work, we focus on the $1/V_{max}$ estimator \citep{schmidt1968, felten1976}, given generically by
\begin{equation}
    \Phi = \sum_i^N \frac{1}{V_{max,i}}
\end{equation}

where $V_{max,i}$ is the maximum possible volume that the $i$th observed galaxy could subtend in a survey detecting $N$ galaxies and still be detected by the given observations. To determine a luminosity function, this estimator can then be evaluated for bins of luminosity, but since the generic form is non-parametric, it could be evaluated as a function of other galaxy characteristics as well.

The fractional completeness of galaxy detections is known as the completeness function, or selection function. Differing methods for calculating the selection function can dramatically impact the resulting corrected LAE distribution. We focus on the $1/V_{max}$ method because its non-parametric form for the LF is easily adaptable to include a selection function dependent on intrinsic luminosity and redshift. However, \citet{herenz2019} showed that in addition to the intrinsic line luminosity of the LAE, the shape of the Ly$\alpha$ halo (LAH) profile may impact the detectability and therefore measurement completeness of an LAE subpopulation. 

Given its complicated escape pathways, Ly$\alpha$ emission is typically not observed tracing the relatively compact stellar population of its host galaxy \citep{malhotra2012}, but is instead found in a spatially extended LAHs \citep{matsuda2012, momose2014, wisotzki2016, leclercq2017, kusakabe2022, guo2023a, herrero2023b}. The Ly$\alpha$ photon's radiative transfer properties also often create broad, complicated spectroscopic profiles \citep{verhamme2006, erb2018, claeyssens2019, blaizot2023, erb2023}, which may even feature two separate ``blue'' and ``red'' peaks, where the observed emission is shifted to lower and higher wavelengths than the Ly$\alpha$ line center predicted by the galaxy's systemic redshift. Integral field unit (IFU) spectroscopy has proven very successful as a method for probing this 3D spatial and spectral profile, especially with observations from the MUSE instrument on the VLT \citep{bacon2010, bacon2017, inami2017, bacon2023} (Vitte et al., submitted; Claeyssens et al., submitted).

At a given redshift and intrinsic line luminosity, the exact spectral shape and spatial distribution of the LAH could significantly affect the rate of LAE detection by effectively spreading the same level of Ly$\alpha$ emission across a broader area, reducing surface brightness and detected signal-to-noise ($S/N$). Furthermore, since Ly$\alpha$ escape is determined at least in part by physical properties of the CGM \citep[e.g.][]{gronke2015, li2022a, li2022b, blaizot2023}, the distribution of LAH properties may hold valuable information on the typical CGM conditions in different eras of cosmic galaxy evolution. Therefore, the advantage gained by studying the properties and distributions of LAHs is twofold: first, by improving standard techniques for correcting incompleteness in LAE observations, the analysis of LAE populations at high-$z$ can be made more accurate; and second, by uncovering the intrinsic distributions of physical parameters of LAHs at a given redshift range, we may develop insights into the nature of the CGM, Ly$\alpha$ escape, and other key questions of galaxy evolution.

In this work, we develop a 3D spatial-spectral LAH model that may be used to replicate the observation of an LAH with given halo properties in specific survey conditions. With this model, we may test the relative importance of different halo characteristics to LAH detectability, and thereby produce a grid of models representing the range of LAH selection functions in a given survey. We then apply this to a sample of $3<z<5$ LAEs from \citet[][hereafter L17]{leclercq2017} in the UDF-Mosaic survey using VLT/MUSE. We use our selection functions and the $1/V_{max}$ completeness estimator to recover intrinsic distributions of the most important LAH parameters, and discuss their implications for further LAH observations and studies.

The paper is organized as follows. In \S2, we describe our methods for construction of the LAH model and for deriving a model's expected detectability. In \S3, we derive a generalized model for LAH selection functions and apply it to an observed sample from L17. In \S4, we recover the intrinsic parameters distributions for $3<z<5$ LAHs and discuss their physical implications and future lines of study. We summarize our results in \S5.

In this work, we use CGS flux units, physical distances, and assume a $\Lambda$CDM cosmology with $\Omega_m=0.3$, $\Omega_{\Lambda}=0.7$, and $H_0=70$ km s$^{-1}$ Mpc$^{-1}$.


\section{Modeling the Lyman-$\alpha$ Halo Selection Function}

\subsection{The Lyman-$\alpha$ Profile}

Determining the selection function for LAHs in a given redshift range and for a given observational setup requires the reproduction of a variety of plausible LAH observations under those conditions. For implementation of this procedure in a MUSE IFU cube, we therefore needed to model the LAH emission in three dimensions, as a miniature datacube consisting of a 2D spatial component and a spectral component covering the extent of the Ly$\alpha$ emission line.

\subsubsection{Spatial Profile}


\begin{figure*}
    \includegraphics[width=\textwidth]{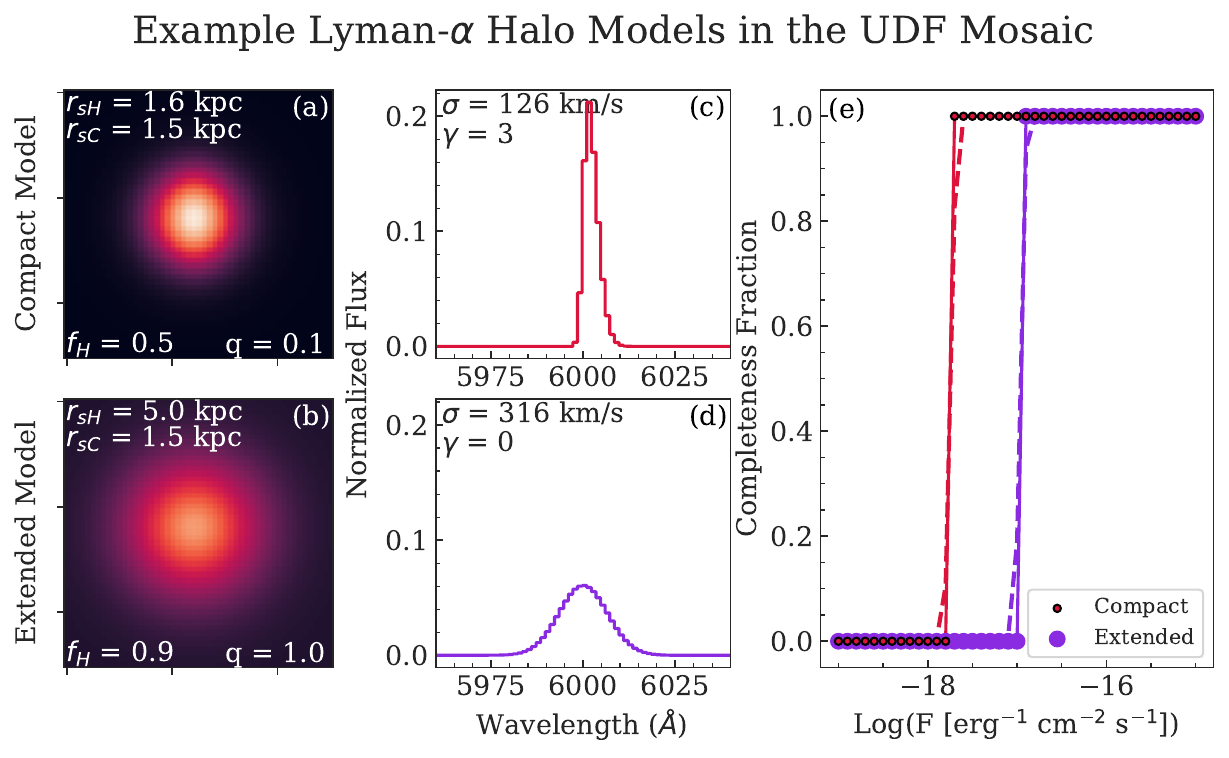}
    \caption{Two example model LAH profiles set in the UDF Mosaic. The left column shows a spatial profile of the model LAH flux, effectively a narrowband image of the Ly$\alpha$ emission, for a model that is spatially relatively compact (panel a) and for a model that is more extended (panel b). The fluxes in each spaxel are log-normalized here to emphasize the distinct shapes. The values of the spatial parameters used for the model are shown in the overlaid text. The middle column gives the normalized Ly$\alpha$ spectral profile for a narrow, skewed line (panel c) and for a broader line with no skew (panel d). The right column (panel e) shows the predicted completeness fraction for the compact and extended model LAHs when inserted into UDF-Mosaic observing conditions at different intrinsic fluxes and at a fixed line center of $\lambda_0=6000\AA$. For a range of fluxes from $Log(F_{Ly\alpha}) = -15$ to $-19$, each model was inserted into a UDF Mosaic datacube and its $S/N$ measured. For a chosen detection threshold $S/N_{det} = 5$, we then estimated the completeness fraction for each insertion. This is given by the red (compact) and purple (extended) circles, and the trend tracked by the solid lines. Dashed lines show error function models of the detectability function, described in \S\ref{sec:sn}. Unsurprisingly, the completeness fraction drops rapidly when the intrinsic flux gets low enough to bring the measured $S/N$ below the detection threshold. This behavior can be accurately modelled by an error function, as shown by the dashed lines in the figure. This panel also demonstrates the clear effect of the LAH parameters on the expected detectability of the LAE: the spatially compact, narrow-line model is detectable at fluxes almost an order of magnitude fainter than the extended, broad-line model.}
    \label{fig:halo_mods}
\end{figure*}

The LAH spatial profile is commonly modelled as a two-component exponential disk \citep{wisotzki2016}, with one component representing a compact, continuum-like central source, and the second representing a diffuse, extended halo. Previous MUSE observations of LAEs have found this to be an effective model \citep[e.g.][]{leclercq2017}, and more recent, high-spatial-resolution studies of lower redshift LAEs have confirmed that more simplified spatial models are insufficient \citep{runnholm2023}. We therefore described the combined profile with:

\begin{equation}
    F(r) = \Sigma_C(r) + \Sigma_H(r) = \Sigma_{C0} e^{-r/r_{sC}} + \Sigma_{H0} e^{-r/r_{sH}}
\end{equation}

where $\Sigma_C(r)$ ($\Sigma_H(r)$) describes the radial flux profile of the compact (halo) component, with a total integrated flux of $F_C$ ($F_H$), and which we model as an exponential disk. The central flux surface brightnesses of the compact and halo components are given by $\Sigma_{C0}$ and $\Sigma_{H0}$, and $r_{sC}$ and $r_{sH}$ are their exponential scale lengths. We define the halo flux fraction $f_H = F_H / (F_H + F_C)$ in order to describe the relative contribution of each component to the total Ly$\alpha$ flux. For examples of two spatial profiles with variations on these parameters, see the panels (a) and (b) of Figure \ref{fig:halo_mods}.

The spatial distribution of the total flux in each component will also depend on its ellipticity, which we measure through the axis ratio $q$. We allow $q$ to vary for the compact component.\footnote{The axis ratio $q$ is related to the other parameters such that $F_{C} \propto r_{sC}^2 \Sigma_{C0} q$, see \citet{peng2010}.}. The ellipticity of the halo component ($q_H$) is fixed to 1. This is a simplifying assumption that has generally been used for large populations of LAHs observed at high redshift \citep{wisotzki2016, leclercq2017}, where low signal makes proper fitting of the halo ellipticity very difficult, but it will not always be the case for individual LAHs, particularly those exhibiting mergers or substantial gas outflows \citep{pessa2024}, such as those driven by quasars \citep{borisova2016, arrigoni2019}. We comment on the potential effects of variable $q_H$ and the assumption of $q_H=1$ in \S\ref{sec:partest}.

\subsubsection{Spectral Profile}

\begin{figure}
    \centering
    \begin{tabular}{c}
    \includegraphics[width=0.5\textwidth]{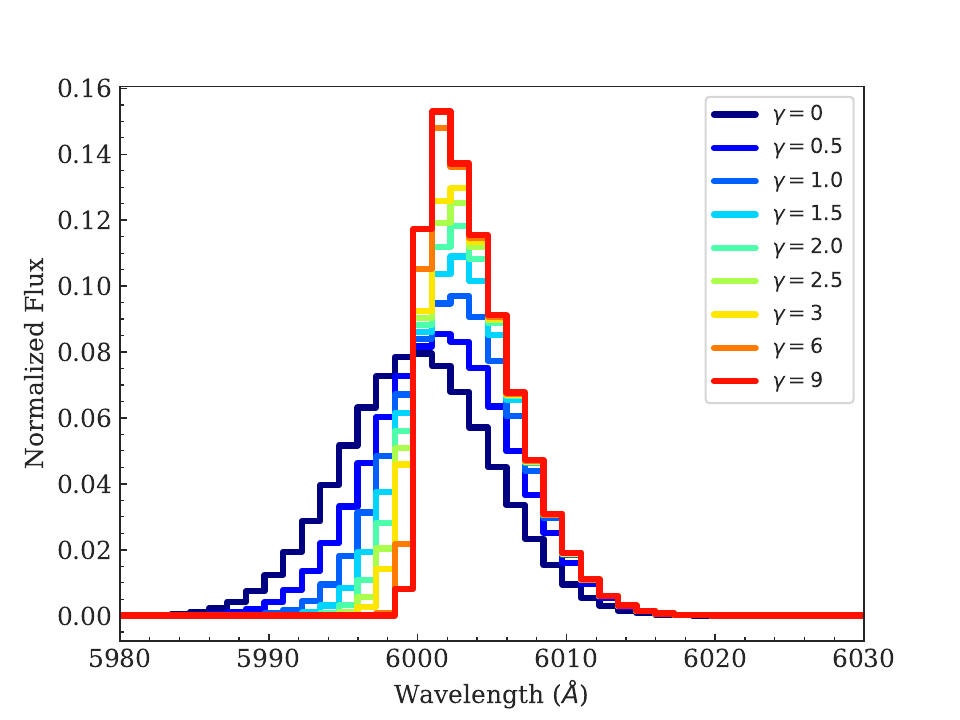} \\
    \includegraphics[width=0.5\textwidth]{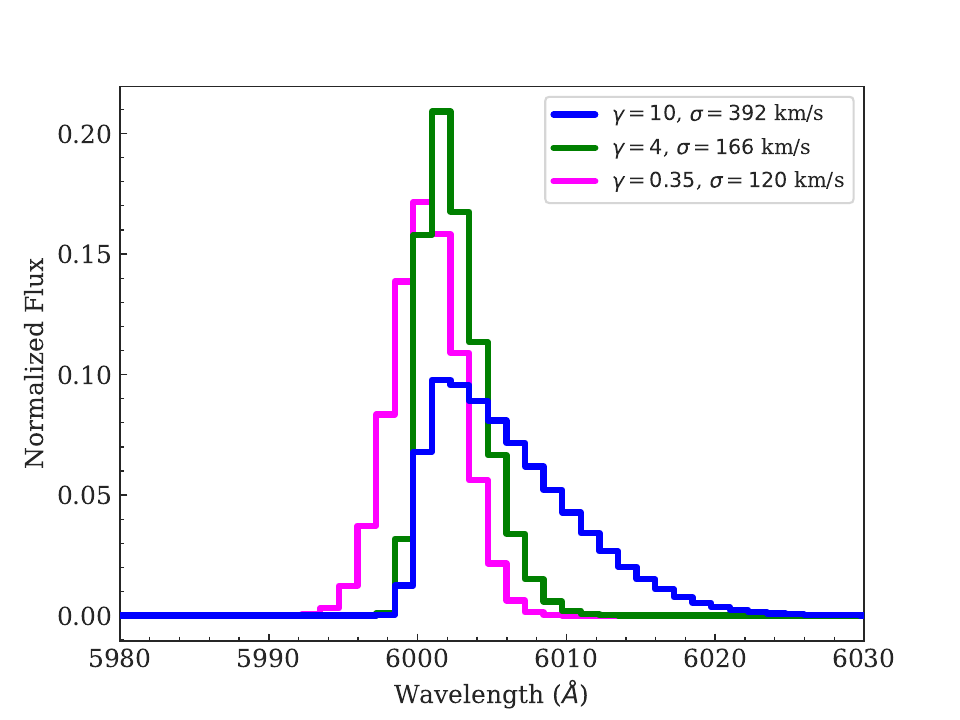}
    \end{tabular}
    \caption{\textit{Top:} example model Ly$\alpha$ profiles at arbitrary and fixed line center $\lambda_0=6000$ \AA, normalized total line flux, and with constant line width $\sigma=250$ km/s. Each color shows the profile with a different skewness factor $\gamma$. The models are resampled to the spectral resolution of VLT/MUSE. \textit{Bottom:} Three normalized Ly$\alpha$ line profiles using the best-fit values of $\sigma$ and $\gamma$ for real observed galaxies from B23, demonstrating some of the variety of shapes of observed Ly$\alpha$ spectral profiles.}
    \label{fig:lya_profiles}
\end{figure}

\begin{figure*}
    \includegraphics[width=\textwidth]{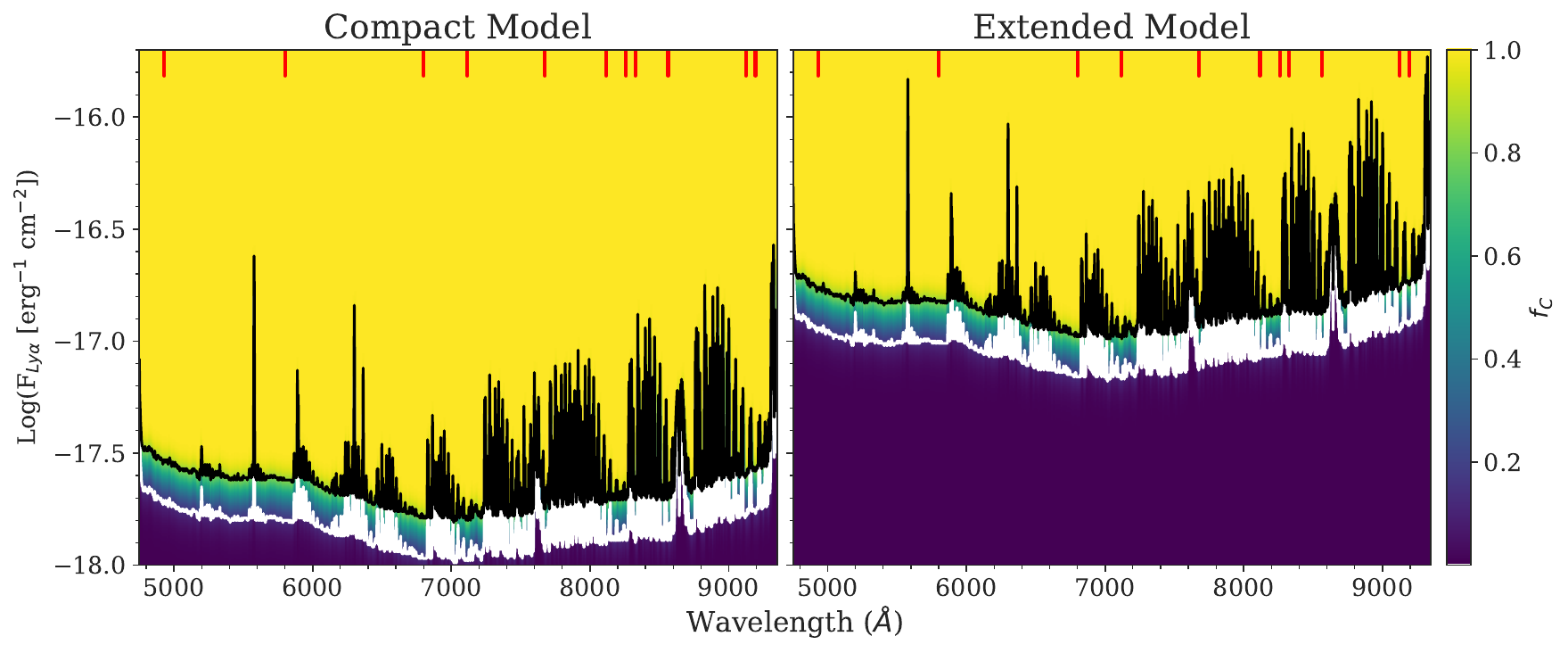}
    \caption{Selection functions for the compact (left) and extended (right) model emitters shown in Figure \ref{fig:halo_mods}, with the color of each flux-wavelength point on the grid giving $f_C$, the detection completeness fraction estimated by the model for the UDF Mosaic. The black and white curves show lines of $f_C = 0.85$ and $f_C=0.15$. The model in the left column represents a relatively compact LAH, with a halo flux fraction $f_H=0.5$, a halo scale length of $r_{sH}=1.6$ kpc, a spectral line width of $126$ km/s, and a skew parameter $\gamma=3$. The model in the right column is more extended, with a halo flux fraction $f_H=0.9$, a halo scale length of $r_{sH}=5.0$ kpc, a spectral line width of $316$ km/s, and a skew parameter $\gamma=0$. Vertical red lines near the top of the panels indicate the wavelengths where models were inserted to LSDCat. The selection functions predict that these changes will have a substantial effect on the detectability, with the more compact model achieving $f_C=1$ down to much fainter line fluxes.}
    \label{fig:sf_model}
\end{figure*}

The scattering nature of Lyman-$\alpha$ emission often creates irregular, sometimes double-peaked spectral line shapes, which may not be adequately replicated with simple Gaussian profiles \citep{shibuya2014}. Since we were primarily concerned with detectability of the line, we focused on modeling the stronger (and usually redder) peak of Ly$\alpha$ emission \citep{laursen2011}. Claeyssens et al. (2024, in prep) find that even when blue peaks are expected to be present, they are unlikely to be detected with MUSE spectral resolution at high redshift. They demonstrate that for a low-$z$ population with blue peak detection rate of 60\% in HST/COS spectra, the rate would drop below 20\% with MUSE observations at $z>3$, so the blue peak is unlikely to be a strong contributor to the expected observed profiles. Previous MUSE observations and related simulations of Ly$\alpha$ emission suggest the blue-peak-dominated fraction of any sample should be small \citep{hayes2021, kerutt2022}.

Some recent MUSE observations and simulations of LAEs do find significant if small fractions with either blue-peak dominated or double-peaked spectral profiles \citep[][Vitte et al. submitted]{blaizot2023, mukherjee2023}. From a detectability standpoint, LAEs with a single dominating blue peak will behave functionally very similarly to those with a single red peak, since our spectral model depends on the observed central wavelength of the dominant peak, and not the systemic redshift, for defining the line's position in the wavelength grid. Ly$\alpha$ spectra with blue and red peaks of comparable flux present a potentially more interesting case, but the inclusion of a second spectral peak would significantly increase the model parameter space to account for a likely small fraction of LAEs. We thus focus on the single-peak case, leaving double peaks to be explored in subsequent work. 

The potentially asymmetric profile of a single peak has been modelled as an asymmetric Gaussian \citep{shibuya2014}, a form used in previous MUSE surveys such as MUSE-WIDE \citep[e.g.][]{herenz2017b}, but more recent MUSE observations have fit Ly$\alpha$ profiles with a skewed Gaussian function \citep[e.g.,][hereafter B23]{bacon2023}, which we adopt as well for consistency with the observed halo parameters we will use. The functional form of the profile is given by:

\begin{equation} \label{eq:skgauss}
    F(\lambda) = \frac{A}{\sigma \sqrt{2\pi}} e^{-(\lambda-\lambda_0)^2/(2\sigma^2)} \times \left( 1 + \text{erf} \left( \frac{\gamma (\lambda-\lambda_0)}{\sigma \sqrt{2}} \right) \right)
\end{equation}

Here the critical profile shape parameters are the line width $\sigma$, typically measured in km/s, and the skew factor $\gamma$, a dimensionless parameter representing the asymmetry of the line profile. A profile with $\gamma=0$ reduces to a symmetric Gaussian, and as the value of $\gamma$ increases, the skewness term becomes more dominant, producing a more asymmetric profile. The central wavelength of the profile is given by $\lambda_0$. Panels (c) and (d) of Figure \ref{fig:halo_mods} show the spectral profiles for two specific LAH models, one narrow and heavily skewed, and the other broad and symmetric. For more examples of the range of variations in model spectra shapes based on the spectral parameters, see Figure \ref{fig:lya_profiles}.

\subsubsection{Modeling the Profile} \label{sec:modprof}

The spatial and spectral components of the profile combined to contribute six total variable parameters to the LAH model: $f_H$, $r_{sC}$, $r_{sH}$, $q$, $\sigma$, and $\gamma$. We also varied two line parameters, the intrinsic total LAE flux $\log_{10}F_{Ly\alpha}$ and the central line wavelength $\lambda_0$. 

For a given set of values of these parameters, we independently modelled the spatial and spectral components of the LAH. First, we used Galfit \citep{peng2002, peng2010} to convolve the two-component disk profile with a Moffat function \citep{moffat1969} representing the point-spread function (PSF) of the observations, using the model spatial parameters and intrinsic line flux. \citet{bacon2017} measured Moffat profile PSF parameters for the UDF Mosaic, which we have adopted here. We modelled the spatial distribution with the 0.2'' pixel$^{-1}$ spatial resolution of MUSE, which for the typical halo length scales at $3<z<5$ fit well in a 51x51-pixel spatial array. For each model, we produced such an array containing effectively a model narrowband image of the LAH.

We modelled the spectral profile in the observed frame according to the skewed Gaussian form described in Equation \ref{eq:skgauss}, and convolved this profile with the expected broadening of the instrumental line spread function (LSF). \citet{bacon2017} determined the LSF through a single-Gaussian model for the UDF Mosaic, such that the FWHM in \AA\ is given by

\begin{equation}
    FWHM (\lambda) = \left( 5.835 \times 10^{-8} \right) \lambda^2 - \left( 9.080 \times 10^{-4} \right) \lambda + 5.983.
\end{equation}

The FWHM evaluated at a given wavelength can then be converted to a Gaussian width and added in quadrature with the ``physical'' line width $\sigma_0$ such that $\sigma = \sqrt{\sigma_0^2 + \sigma_{LSF}^2}$, where $\sigma$ is the observed line width described in Equation \ref{eq:skgauss}.

We then multiplied each pixel in the 2D spatial profile by the normalized skewed Gaussian spectral profile with given $\lambda_0$, $\sigma$, and $\gamma$, distributing the line flux fully in a 3D minicube model of the LAH across the full wavelength MUSE coverage. Here we assumed a constant spectral profile shape across the entirety of the spatial profile, which is not necessarily the case for observed LAHs \citep{claeyssens2019, guo2023b, mukherjee2023}. However, these observed spectral variations occur on scales small enough to require substantial stacking or gravitational lensing to confidently observe at high redshift, and may thus be expected to have a small contribution relative to the predominant halo characteristics. As such, the spectral component of our model may be treated as primarily representing the brighter, inner regions of the halo which will dominate the detectability, and the model uses the same values for $\sigma$ and $\gamma$ for all spaxels. We then combined this minicube with the variance spectrum and exposure map from the related observations to fully simulate the observing conditions of the model LAH.

\subsection{Signal-to-Noise Estimation and Measuring the Completeness Fraction} \label{sec:sn}

To evaluate the detectability of a given LAH emission profile, we used the Line Source Detection and Cataloguing Tool \citep[LSDCat,][]{herenz2017a, herenz2023}, a 3D matched-filtering package for emission line detection developed for IFU observations. LSDCat operates by cross-correlating a 3D emission line template with a continuum-subtracted IFU datacube. For the line template, we use a 3D Gaussian with a spatial width adjusted to the wavelength-dependent PSF, and with a spectral width of 250 km/s. These template choices have been optimized for Ly$\alpha$ searches with LSDCat and have demonstrated success in previous MUSE surveys \citep[e.g. in MUSE-WIDE,][]{herenz2019, urrutia2019}. After cross-correlation with the cube and its associated variance spectrum, LSDCat returns a signal-to-noise ratio ($S/N$) cube.

We initially tested the source recovery of the models by inserting LAH models into 51x51-pixel UDF-Mosaic-like minicubes at varying levels of intrinsic flux. We ran LSDCat on each minicube and recorded the measured $S/N$ for that model and intrinsic line flux. Setting a threshold level of $S/N$ for detection at $S/N_{det} = 5$, a common observational cutoff for significant detections, we then evaluated the expected completeness fraction $f_C$ for each inserted model. This can be seen in panel (e) of Figure \ref{fig:halo_mods}, which shows the expected completeness fraction for the example extended and compact halo models as a function of the intrinsic line flux at a fixed line center of $\lambda_0=6000$ \AA. Unsurprisingly, for very bright fluxes, the measured $S/N$ is substantially higher than the threshold, and the expected detectability fraction is essentially 1. At very faint fluxes, the $S/N$ is well below the threshold, and the recovery is essentially 0. In this simple scenario, where we insert and measure just a single LAH at each flux, the detectability function transitions rapidly from one stage to the other as the intrinsic flux of the models dims and $S/N \rightarrow S/N_{det}$, with a narrow regime of $0<f_C<1$.

This functional behavior of the completeness fraction has been observed in tests from \citet{herenz2019} as well, wherein models representing real observed LAHs were inserted and recovered in MUSE-WIDE observing conditions. They also performed multiple insertion tests for each flux level, and so were able to account for variation in the $S/N$ measurement from noise fluctuations across the observed field of view. Since this form is reproduced in both observation-based and analytical models, it is safe to then model the completeness fraction as a function of flux. In the simplest test case we performed above, the transition from $f_C=1$ to $f_C=0$ is very narrow, but this assumes a perfect measurement of the $S/N$ ratio, whereas repeated insertion tests of the same models produce a larger spread in the measured recovery function. If the error in measuring the $S/N$ compared to $S/N_{det}$ is assumed to be Gaussian, then the behavior of the resulting detectability function accounting for the normal distribution of error in the $S/N$ may be described by an error function\footnote{The error function is defined such that $\text{erf}(z) = \frac{2}{\sqrt{\pi}} \int_0^z \exp^{-t^2} dt$}. Dashed lines in panel (e) in Figure \ref{fig:halo_mods} show that this representation closely matches the full-insertion tests. 

\subsection{Systematic LSDCat Insertion and Line Recovery}

Given that the full-model-insertion method described in \S\ref{sec:sn} is well-described by an error function, it became possible to replace the time-intensive insertion of the model at every level of intrinsic flux with a more efficient approach: insert the model at a single intrinsic flux, measure the $S/N$ ratio at that level, and then model the $S/N$, and therefore completeness fraction, for a whole range of possible observable fluxes. This approach is much more practical for attempting to model completeness for a whole LAH population, which may consist of LAHs with a wide range of combinations of halo parameters.

Similarly, it was not necessary to test each model at every possible central wavelength observable by MUSE. With the exception of changing effects of the LSF, the signal measured for a model with a fixed line flux should not change with wavelength, and the LSF effect is generally small relative to the physical line widths, and changes slowly with wavelength. Therefore, the main changes to the $S/N$ measure with wavelength should be expected to depend on the noise properties of the MUSE observations, which has two main drivers: the MUSE sensitivity curve, and the presence of atmospheric skylines. The $S/N$ needed to be measured at a set of wavelengths that accurately accounted for these two factors. 

We selected nine separate wavelengths for model LAH insertions: 4930, 5800, 6800, 7116, 7674, 8260, 8563, 9125, and 9193 \AA. These wavelengths were chosen to provide a sufficient sampling of the underlying MUSE sensitivity curve (and thus variance spectrum) in relatively skyline-free spectral regions. After sampling the $S/N$ at each of these wavelengths with LSDCat, we interpolated a full $S/N$ curve as a function of wavelength at fixed line flux by using the known variance spectrum for the MUSE UDF Mosaic. This incorporated both the shape of the underlying MUSE sensitivity and the impact of skylines. Then at each wavelength, we modelled $f_C$ for a range of intrinsic line fluxes.


This process yielded a selection function for a specific LAH spatial-spectral profile as a function of the intrinsic line flux and the central line wavelength. Examples of this 2D selection function are shown in Figure \ref{fig:sf_model} for the compact and extended model LAHs depicted in Figure \ref{fig:halo_mods}. The selection functions are grids of intrinsic line flux and wavelength, with the value at each grid coordinate the $f_C$ expected for that model at the specific flux and wavelength (shown by the color in Figure \ref{fig:sf_model}). Each column in the selection function is the same function as is depicted in panel (e) of Figure \ref{fig:halo_mods}. This immediately demonstrates the impact that variation in the LAH parameters has on Ly$\alpha$ detectability: the compact LAH with a narrow spectral profile is expected to be complete down to almost an order of magnitude fainter intrinsic flux compared to the extended, broad-line halo. To gain a complete understanding of the distribution of LAH properties, it therefore becomes necessary to model this selection function across the range of possible LAH parameters.

\section{A General LAH selection function}

\subsection{Parameter Tests} \label{sec:partest}

Determining completeness calculations for the range of possible observed LAHs necessitates a multi-dimensional grid of selection functions, accounting for variation in each of the six model parameters as well as intrinsic line flux and central wavelength (redshift). Such a grid is potentially very computationally expensive to both generate and analyze. It was thus helpful to consider the relative impact each parameter may have on changes in LAH detectability, such that we could coarsen/refine the grid resolution in that parameter space accordingly. 

\begin{figure*}[]
    
    \centering
    \begin{tabular}{c}
         \includegraphics[width=0.85\textwidth]{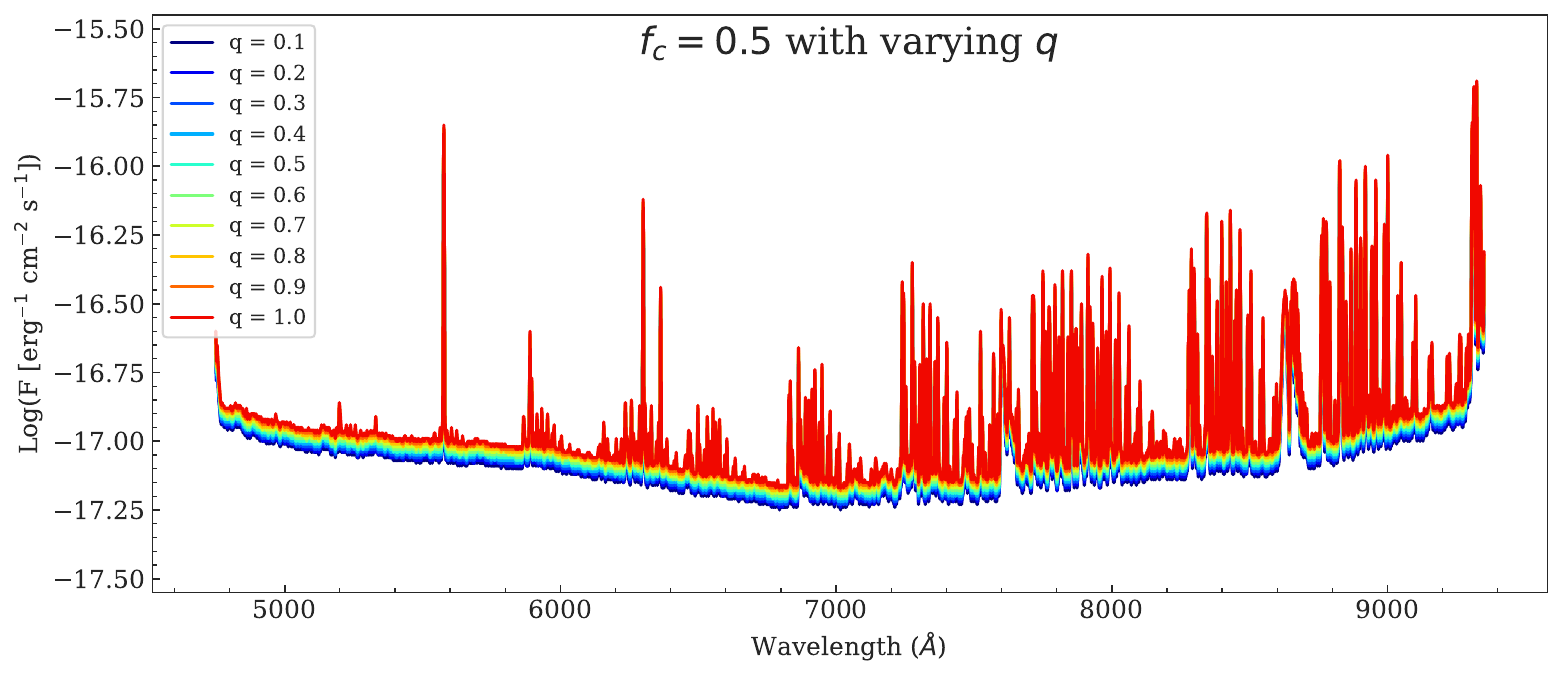} \\
         \includegraphics[width=0.85\textwidth]{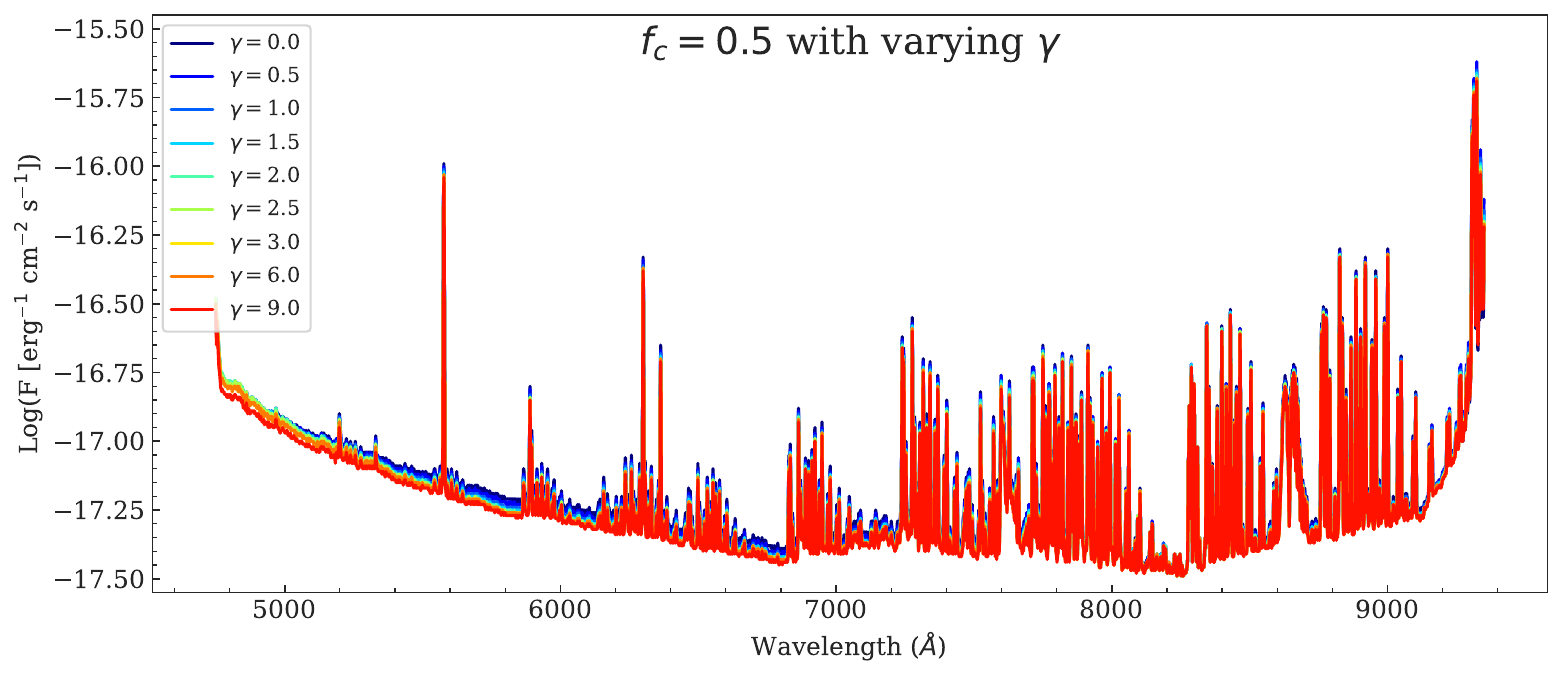} \\ 
         \includegraphics[width=0.85\textwidth]{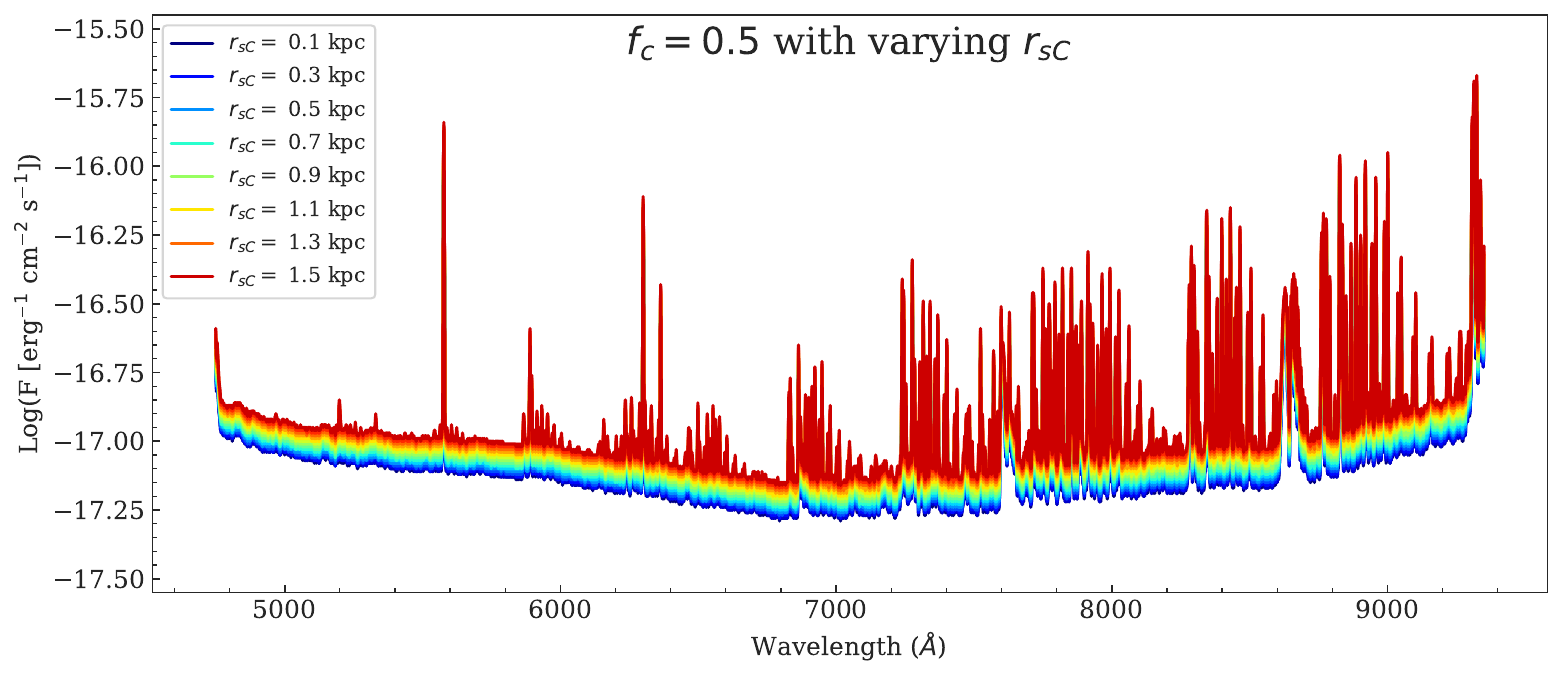}
    \end{tabular}
    \caption{Individual parameter tests for three of the LAH model parameters. Each panel gives the completeness fraction $f_C=0.5$ as a function of intrinsic line flux and wavelength for a commonly observed range of a given parameter's values (shown by the colored curves), with the remaining 5 parameters held fixed at an average value. This demonstrates the variable contribution to the detectability each parameter may make, with $\gamma$ changing the 50\% detection point by only $\sim$0.05 dex in intrinsic flux, while the $q$ and $r_{sC}$ lead to average changes of 0.08 and 0.13 dex, respectively.}
    \label{fig:param_tests}
\end{figure*}

\begin{figure*}[]

\centering
\begin{tabular}{c}
    \includegraphics[width=0.85\textwidth]{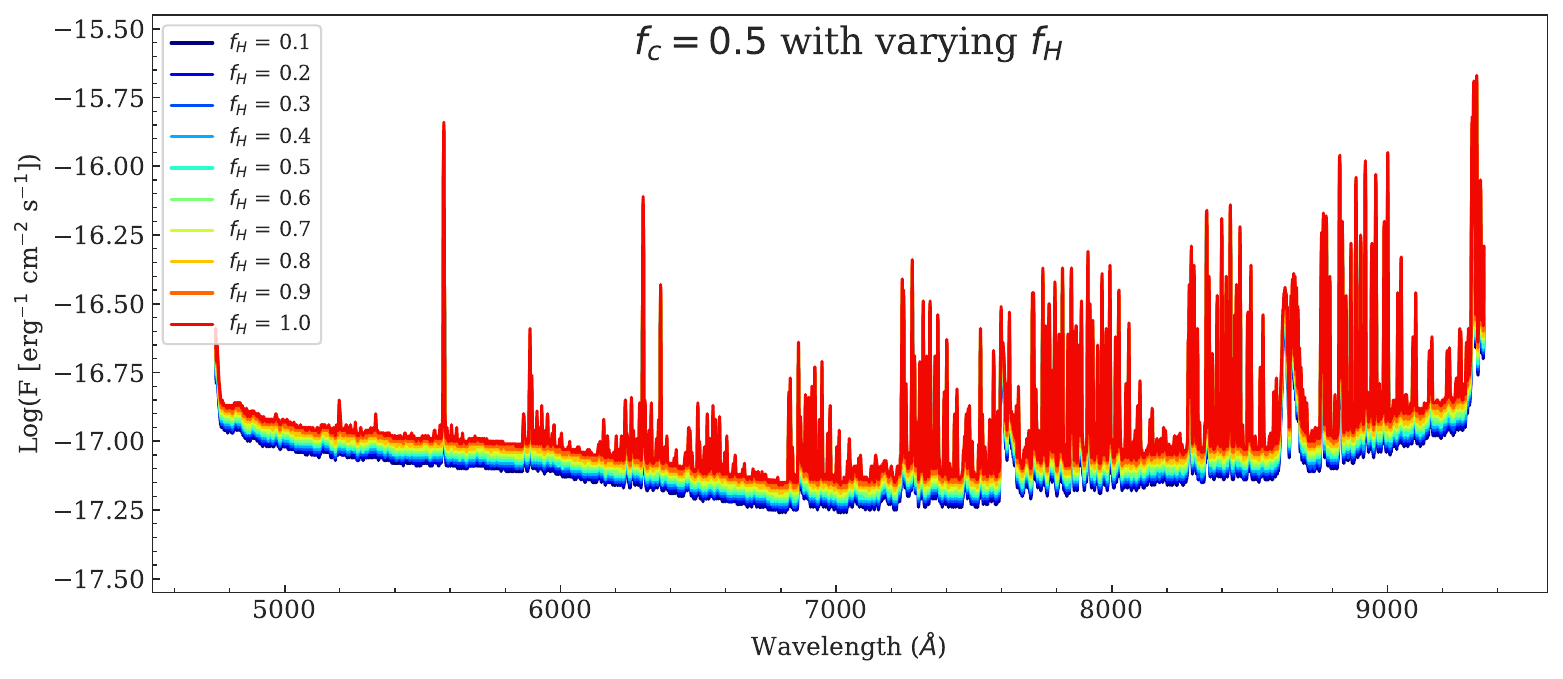} \\
    \includegraphics[width=0.85\textwidth]{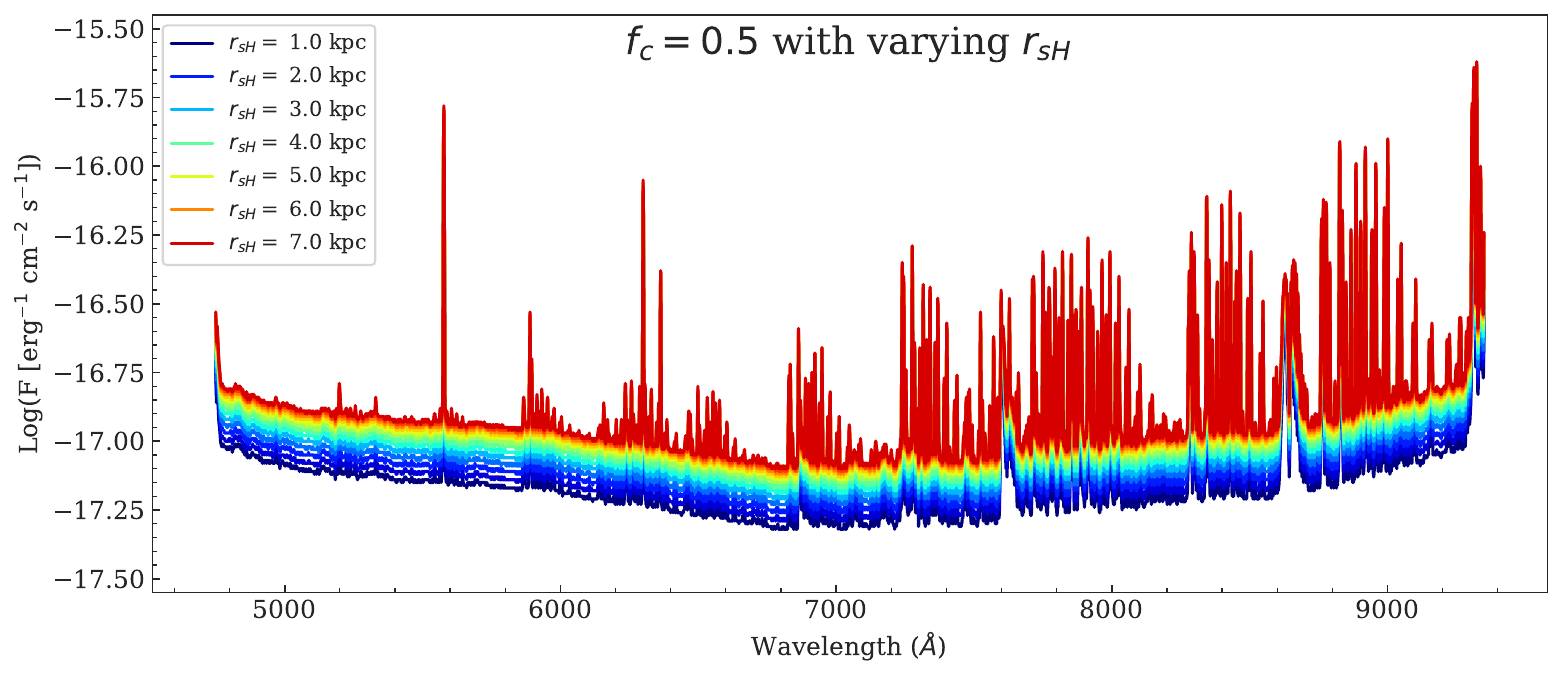} \\
    \includegraphics[width=0.85\textwidth]{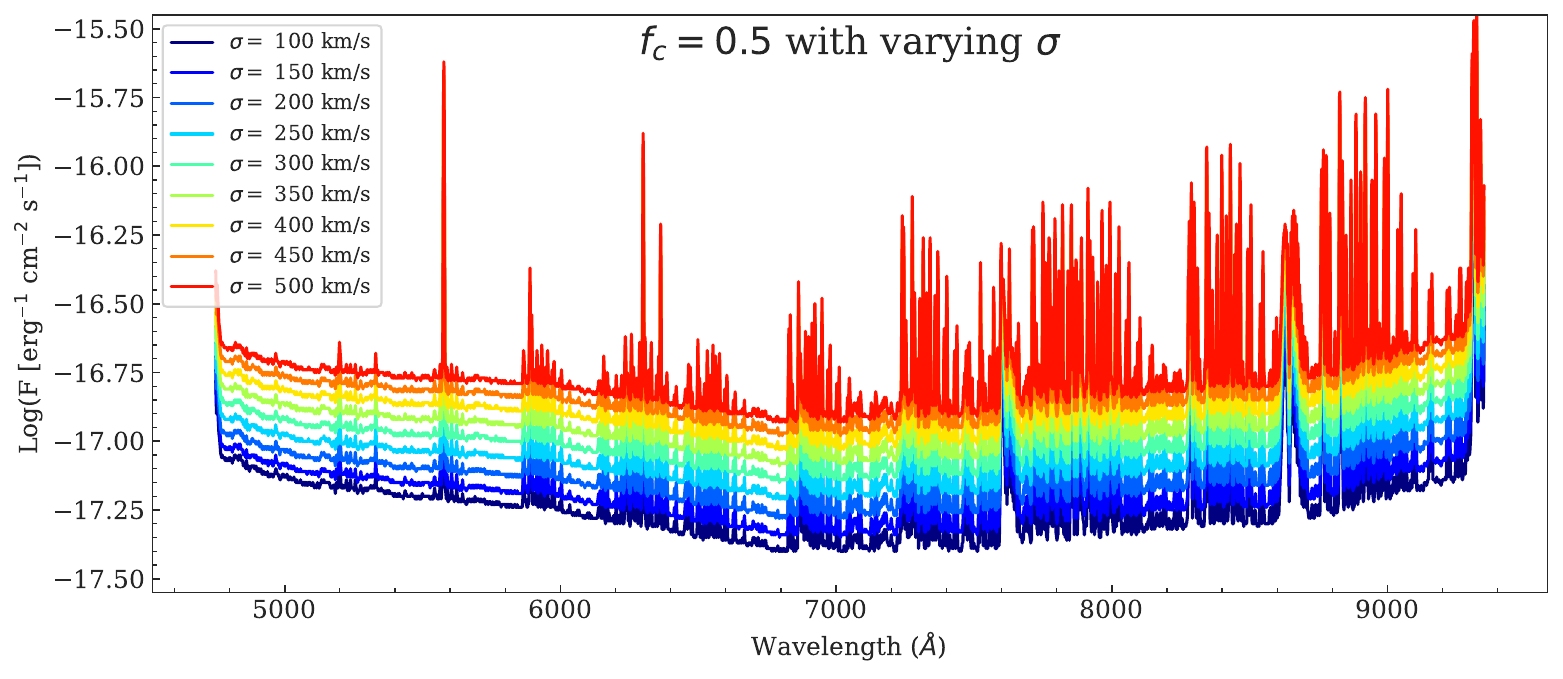}
    
\end{tabular}
\caption{As in Figure \ref{fig:param_tests}, but with the three remaining LAH model parameters. These parameters show much more substantial influence on the detectability than the previous set, with $f_H$ producing changes of $\sim$0.12 dex, $r_{sH}$ changing by $\sim$0.22 dex, and $\sigma$ generating changes in the $f_C=0.5$ point of up to 0.5 dex.}
\label{fig:param_tests2}
\end{figure*}

We measured this by fixing five of the six LAH parameters to an average value from observed LAHs in MUSE surveys (described in detail in \S\ref{sec:udf}), then running the LSDCat recovery test on the observed range of values for the sixth parameter. We tested this in the CANDELS-CDFS-03 field from MUSE-WIDE \citep{urrutia2019}, a relatively shallow field with well-measured noise properties. The test results are shown in Figures \ref{fig:param_tests} and \ref{fig:param_tests2}. We plot the flux at which the modelled completeness fraction $f_C=50\%$ as a function of wavelength, yielding a separate ``50\% Curve'' for each value of the parameter being tested, shown by different colors. Because the selection function closely resembles an error function, which rapidly changes from 1 to 0, changes in the $f_C=0.5$ point ($F_{50}$) will be particularly sensitive to the detectability effects of the changing parameter, making it a useful diagnostic for this test. 

We found the most impactful parameters to be the line width $\sigma$ and the halo scale length $r_{sH}$ (both in Figure \ref{fig:param_tests2}). Changes in $\sigma$ moved the $F_{50}$ flux by up to 0.5 dex in intrinsic line flux at a given wavelength, and variation in $r_{sH}$ could move $F_{50}$ by over 0.2 dex. The halo fraction $f_H$ and the compact-component scale length $r_{sC}$ had more moderate effects ($\sim0.12-0.13$ dex), and the compact ellipticity $q$ and the skewness factor $\gamma$ produced relatively small changes in isolation ($\leq0.1$ dex). Here we also tested the possible effects of our assumption of circular halo shape. Fixing average values for the six listed parameters, we checked the $F_{50}$ curve for $q_H=1$ and $q_H=0.1$. This maximal variation produced an average shift of $\sim$0.07 dex, comparable to the effects of $\gamma$ and $q$. Given this and the fact that L17, the primary source of our observational constraints, assumed circular halo shapes, we will maintain a fixed $q_H=1$, though we note that more elliptical halos will have slightly improved detectability. These effects may be explored more deeply in subsequent work with an expanded LAH sample. 

We note as well that this test does not account for possible inherent correlation in the LAH parameters, instead treating them as independent. Consequently, the completeness fractions here as a function of flux should not be considered in absolute terms, but only in the magnitude of the relative change in completeness with the variable parameter. However, even if some correlations between parameters are later discovered, there is initially nothing to indicate that any such connection would cause the detectability properties to change from the smooth, monotonic changes observed in these tests. Potential correlations between halo parameters will be explored further in Section \ref{sec:imp}.

\subsection{Observational Sample: The UDF Mosaic}
\label{sec:udf}

The major analysis in this work uses data from the Hubble Ultra Deep Field (UDF) survey mosaic \citep[][]{bacon2023}. The data were observed between September 2014 and February 2016 with the MUSE/VLT instrument as part of the MUSE consortium guaranteed time observations. The mosaic consists of a grid of nine 10-h observations of 1'x1' MUSE pointings in the Hubble Deep Field South (HDFS). Overlapping the mosaic is a single 20-hr pointing, denoted UDF-10, for a cumulative 30-hr of deeper exposure. We note here that though UDF-10 observations contribute to the observed distribution of LAH parameters, the model results in this work will only address the conditions of the mosaic. The UDF data reduction is described in detail in \citet{bacon2017}, and produced a mosaic datacube with a wavelength range of $3750 < \lambda < 9350$, and average resolution of $R\sim3000$, and a spatial resolution of 0.2"x0.2" per pixel. The PSF and LSF characteristics of the observations are well-studied, as described in \S\ref{sec:modprof}. 

\clearpage

\clearpage

\begin{figure*}[h!]
    \includegraphics[width=\textwidth]{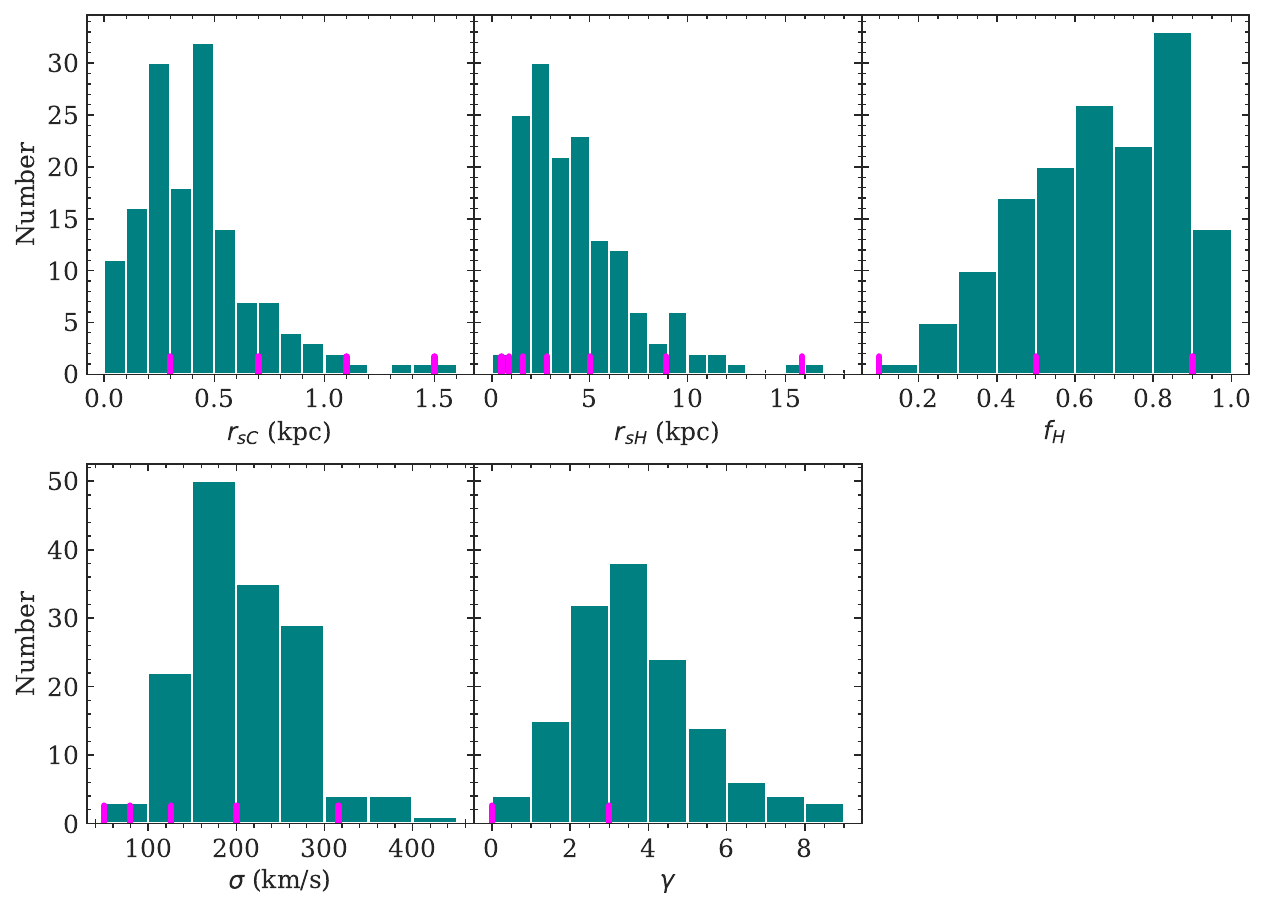}
    \caption{\textit{Top:} The observed distributions of the compact scale length, halo scale length, and halo fraction in LAHs from the L17 catalog. \textit{Bottom:} The observed distributions of the line widths and skewness factors for the same sample of LAHs, using the spectral fit results from the B23 catalog. Vertical magenta lines indicate the grid sampling values described in Table \ref{t_udf_grid}.}
    \label{fig:hists}
\end{figure*}

This dataset is convenient for our purposes because the depth of observation should be able to probe a range for each LAH parameter, necessary for a comprehensive analysis of the detectability. L17 performed Ly$\alpha$ halo measurements for 145 continuum-faint galaxies in the UDF Mosaic and UDF-10, providing a catalog of $r_{sH}$, $r_{sC}$, and halo- and compact-flux measurements from which a halo fraction could be determined. Spectral analysis in B23 provided measures for $\sigma$ and $\gamma$. Distributions of these parameters in the observed sample are shown in Figure \ref{fig:hists}. Ellipticity measurements are not commonly available for the L17 sample, especially since many of the continuum-faint sample do not have resolved compact components. In the Lensed Lyman-Alpha MUSE Arcs Sample (LLAMAS), \citet{claeyssens2022} took advantage of the magnification from gravitational lensing to measure the ellipticity distribution for a comparable LAE sample \citep{richard2021}. They found half the LAEs consistent with circular compact components ($q=1$), with the other half relatively evenly distributed across other values of $q$. As the lensing magnification provides both improved spatial resolution and access to fainter sources, it is reasonable to use this as a statistically comparable distribution for the ellipticities. 

Prior to fitting the LAH parameters, the L17 sample went through several selection cuts, as described in \S2.2 of that paper. Most selection criteria were for data quality purposes and are unlikely to bias the sample toward or against a particular configuration of halo parameters. The one caveat to note is the removal of LAEs in close pairs (defined as having $<$50 kpc projected transverse separation and $<$1000 km/s velocity offsets). It is possible that interacting systems could have systematically different distributions in one or more of the halo parameters, but determining this would require a large sample of interacting LAEs with well-fit LAH parameters, which is currently not available. The characteristics of this sub-population will thus be investigated in future work, and the distributions derived in this work will describe the individual LAE population.




\subsection{Selection Function Grids}

With this set of LAHs described in Sec. \ref{sec:udf} as a comparison sample, we designed a parameter grid for which a large, spanning set of selection functions could be generated for LAHs in the UDF Mosaic. The grid design is described in Table \ref{t_udf_grid}, and the grid points are shown relative to the observed sample in Figure \ref{fig:hists}.

With the potential for a six-dimensional grid to consume substantial computing time, we chose to sample the grid with different step patterns for each parameter based on the parameter's expected influence on the selection function as measured in the single-variable tests described in \S \ref{sec:partest}. We selected the two most influential parameters, $\sigma$ and $r_{sH}$, to have finer samplings spanning the observed parameter distributions shown in Figure \ref{fig:hists}. Given the distributions show strong peaks at lower values of $\sigma$ and $r_{sH}$ with long tails at higher values, we opted to sample these parameters in log space, yielding a finer grid where most objects are observed.

We sampled the remaining parameters in linear steps, and with larger step sizes amounting to just 2-3 sample points per parameter axis. We note as well that for $\gamma$, we do not sample the entire observed range, stopping instead at a maximum skewness of $\gamma=3$. As we observed very little change in the selection function for changes in $\gamma$ beyond $\gamma \gtrapprox 2$ (see Figure \ref{fig:param_tests}), it should suffice for the model to simply test the difference between unskewed ($\gamma \approx 0$) and skewed ($\gamma \gtrapprox 2$) profiles.

Generating a grid to the specifications given in Table \ref{t_udf_grid} produced 2352 individual LAH models and associated selection functions. 

\begin{table}
\centering

\caption{Grid parameters for UDF Mosaic}
\begin{tabular}{lcccccc}
\hline\hline
 & $q$ & $\gamma$ & log($r_{sH}$) & $r_{sC}$ & $f_H$ & log($\sigma$)\\
 & & & (kpc) & (kpc) & & (km/s) \\
\hline
 $X_{min}$ & 0.1 & 0 & -0.3 & 0.3 & 0.1 & 1.7 \\
$X_{max}$ & 1 & 3 & 1.2 & 1.5 & 0.9 & 2.9 \\
$\Delta_{step}$ & 0.9 & 3 & 0.25 & 0.4 & 0.4 & 0.2 \\
\hline
$N_{steps}$ & 2 & 2 & 7 & 4 & 3 & 7 \\ 
Cumulative & 2 & 4 & 28 & 112 & 336 & 2352 \\
\hline
\end{tabular}
\label{t_udf_grid}
\end{table}

\subsection{Marginalized Selection Functions} \label{sec:margin}

\begin{figure}
    \centering
    \includegraphics[width=0.5\textwidth]{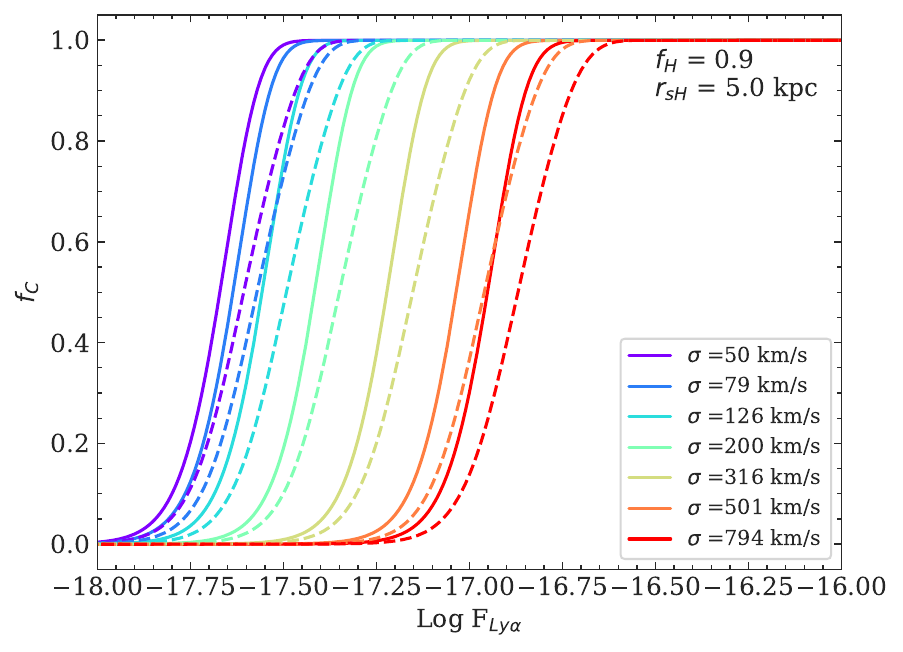}
    \caption{Completeness fraction curves for a series of LAH models that marginalize over the grid axes of the $r_{sC}$, $q$, and $\gamma$ parameters at a fixed central wavelength of $\lambda_0 = 6800$ \AA. These models show fixed values of $f_H$ and $r_{sH}$ but with variable $\sigma$, indicated by the different colors of the curves. The dashed lines show a flat marginalization, where a simple average is taken over the models of variable $r_{sC}$, $q$, and $\gamma$ at the given $f_H$, $r_{sH}$, and $\sigma$ values. The solid lines show the results of a marginalization weighted by the observed distributions of the marginalized parameters (as shown in Figure \ref{fig:hists}). This weighted marginalization increases the contribution of models with smaller $r_{sC}$ lengths, resulting in completeness curves that drop off at fainter fluxes.}
    \label{fig:marginalize}
\end{figure}

The relatively coarse sampling of the selection function grid does pose a potential problem for modeling specific LAH profiles that fall between the grid points. This requires a more generalized adaptation of the selection function grid, whereby the selection function for any potential combination of parameter values within the grid ranges can be individually modelled. 

To reduce the complexity of the models, we chose to marginalize over the less-impactful LAH parameters. Selecting $q$, $\gamma$, and $r_{sC}$ as the parameters with the least expected influence on the selection function\footnote{Though $r_{sC}$ and $f_H$ had similar impacts on detectability in the single-parameter tests, observations of $f_H$ are more likely to span the full tested range, given the rarity of $r_{sC}$ values greater than 1 kpc as measured from high-resolution HST imaging, and the limitations of measuring low-$r_{sC}$ with a seeing-limited PSF. Therefore, we expect the practical effects of $f_H$ to be more substantial.}, we averaged the completeness curves at each wavelength along the axes of these variables for each set of fixed values of $\sigma$, $r_{sH}$, and $f_H$. 

Figure \ref{fig:marginalize} shows a series of example marginalized completeness curves for models with $r_{sH}$ fixed to 5 kpc and $f_H=0.9$, while varying the line width $\sigma$. The marginalization is shown in two possible implementations. First, the dashed curves show flat marginalizations, taking a simple average of completeness curves across models of all values of $q$, $\gamma$, and $r_{sC}$. However, since we have shown that variations in these parameters may still have some influence on the selection function (see Figure \ref{fig:param_tests}), it would likely produce more accurate results for the marginalization to capture as much of the true distribution of those parameters as possible. 

We accomplished this via a weighted marginalization, in which the completeness curve average is weighted by the observed distributions of the marginalized parameters as shown in Figure \ref{fig:hists}. The primary impact of this is to increase the contribution of models with short $r_{sC}$ values ($r_{sC} < 0.5$ kpc), which are detected much more frequently and are expected to have the largest impact on detectability out of the remaining three parameters. We weighted the marginalizations along $q$ by the distribution measured by \citet{claeyssens2022}. For the skewness parameter, we weighted by the $\gamma$ distribution from B23.

The results of this weighted marginalization are shown as solid lines in Figure \ref{fig:marginalize}. The general effect of this marginalization is to shift the completeness curves to fainter fluxes, with the $F_{50}$ point shifting 0.05-0.1 dex lower in $\log_{10}F_{Ly\alpha}$. This makes intuitive sense, given the increased weight to low-$r_{sC}$ models, whose compactness improves their detectability. Of course, weighting the marginalization to observed distributions of the marginalized parameters does potentially induce a bias by underweighting the contributions of less-detectable parameter values (e.g., very extended core scale lengths or very high skewness values). But since we are marginalizing only those parameters previously determined to have small influences on detectability across their entire observed ranges, such a bias is likely to be small. LAHs with very high $r_{sC}$ are not so much less detectable than more compact LAHs that missing some will substantially affect the marginalization weights, for example. Thus, we used this set of marginalized models to reduce the parameter space and explore in more detail the distributions of the remaining three LAH parameters.

\section{Recovering the Intrinsic LAH Distribution}

\subsection{The LAH Sample} \label{sec:sample}

\begin{figure*}
\centering
    \includegraphics[width=0.9\textwidth]{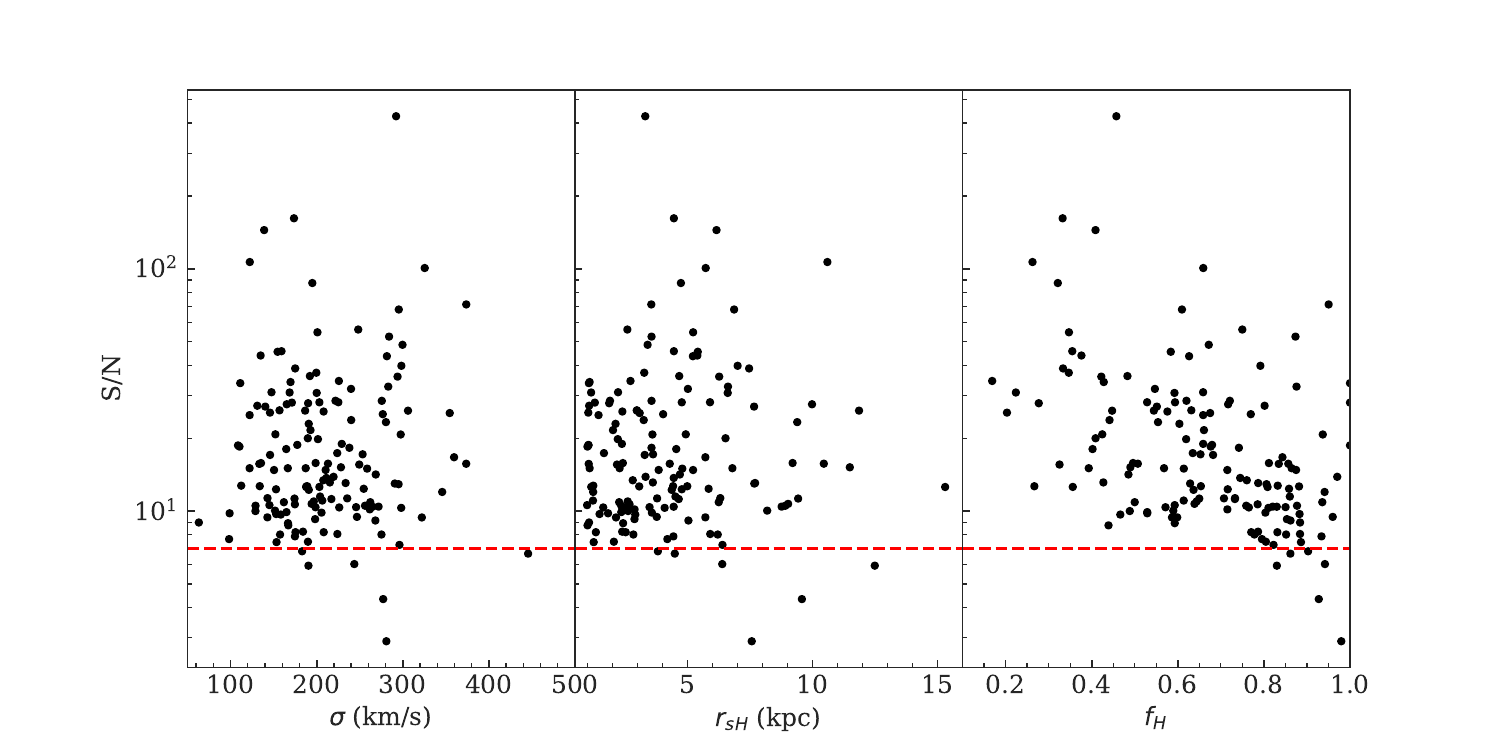}
    \caption{The $S/N$ measured from LSDCat for the L17 LAH sample as a function of three LAH parameters: $\sigma$, $r_{sH}$, and $f_H$. Black dots show the individual LAH values, and the red dashed line shows the adopted SNR cutoff at $S/N > 7$. One feature to note is the $S/N$ behavior of $f_H$, particularly for lower halo fractions. LAHs with low $f_H$ are generally not detected with low $S/N$, despite low halo fractions' advantages in detectability. We discuss this further in \S\ref{sec:sample} and \S\ref{sec:imp}.}
    \label{fig:snr}
\end{figure*}

We used the generalized selection function grid to recover the intrinsic distributions of the line widths, halo scale lengths, and halo flux fractions of $3<z<5$ LAHs. To begin with, we took a subsample cut of the L17 LAH sample at a $S/N$ level where approximately the full range of observed parameters are still observed. Our $f_C$-based completeness correction can correct a low number of detections to a higher, intrinsic number, but such a correction could not be applied to zero detections. Ideally we would make use of as large an observed LAH sample as possible for the best number statistics in assessing the population, but including LAHs at a lower $S/N$ could bias even the corrected distribution, since we could not correct for parameter values that dropped below detection altogether. 


We assessed the proper cutoff via signal-to-noise diagrams, shown in Figure \ref{fig:snr}. Each panel in the figure gives the $S/N$ ratio as a function of $\sigma$, $r_{sH}$, and $f_H$. We needed to select a $S/N$ cut at a level where LAHs were detected across the total observed range of each parameter. For this, we chose $S/N = 7$, indicated by the red dashed line in Figure \ref{fig:snr}. At this approximate level of signal, there is at least one detection even among the rarer LAHs with very high $\sigma$ or $r_{sH}$. 

The one exception is $f_H$, where no low-$S/N$ LAHs are detected for very low halo flux fractions. There is an apparent anticorrelation between the $S/N$ and $f_H$, with only high-fraction LAHs detected lower signal. Given the results of the parameter detectability tests described in Sec. \ref{sec:partest}, which suggest that lower halo flux fraction \textit{increases} detectability, this implies that lower-signal low-$f_H$ galaxies may simply be more intrinsically rare. This would support findings that lower flux fractions are found with higher intrinsic Ly$\alpha$ luminosities at both low and high redshift \citep{hayes2014, ostlin2014, wisotzki2016, leclercq2017}. Given this, we may take the $f_H$ subsample to be complete even if no low-$f_H$ LAHs are detected in the sample at $S/N=7$.

\subsection{The 1/$V_{max}$ Estimator}

The 1/$V_{max}$ method is a non-parametric estimator and commonly used completeness metric derived for measuring a galaxy luminosity function \citep{schmidt1968, felten1976}. As a non-parametric method, there is no underlying assumption of distribution shape, so this estimator may be readily adapted to estimate an unknown distribution of galaxies as a function of halo parameters other than luminosity. A generic form of a binned, differential distribution given by the 1/$V_{max}$ estimator may be shown as:

\begin{equation} \label{eq:dif}
    \phi_{1/V_{max}} (\langle X_k \rangle) = \frac{1}{\Delta X_k} \sum_i \frac{1}{V_{max,i}}
\end{equation}

where $X_k$ is a given LAH physical parameter out of a set of $k$ parameters being measured, and each $V_{max}$ in the sum is evaluated for the $i$th out of $N$ galaxies. In this approach, the 1/$V_{max,i}$ terms are summed over binned values of $X_k$, the bin width for which is given by $\Delta X_k$.

The term $V_{max}$ represents the maximum volume within which a given galaxy could still be detected and included in a given observation \citep{johnston2011}, and so may be used as a completeness estimator. We use a definition for $V_{max}$ modified to account for redshift- and luminosity-dependent detection \citep{caditz2016, herenz2019}. That definition is given by:

\begin{equation} \label{eq:vmax}
    V_{max,i} = \omega \int_{z_{min}}^{z_{max}} f_c(L_{Ly\alpha},z) \frac{dV}{dz} dz
\end{equation}

In this definition, $\omega$ is the angular area subtended by the survey, and $z_{min}$ and $z_{max}$ are the lower and upper bounds of the redshift range under consideration. For the UDF Mosaic, $\omega=9$ arcmin$^2$, corresponding to nine 1'x1' MUSE fields of view. We take $z_{min}=3$ and $z_{max}=5$, corresponding to the redshift range of the L17 sample after our SNR cuts. The $dV/dz$ term is the differential cosmological volume element \citep{hogg1999}.

The term $f_c(L_{Ly\alpha},z)$ represents the selection function, the detection completeness fraction as a function of the intrinsic Ly$\alpha$ line luminosity and the redshift of the observed halo. For each LAH in the L17 sample cut, we used the halo parameters from the L17 and B23 fits to model $f_c$ for a specific LAH. We took the total line fluxes measured in L17 for each LAH, converted to the intrinsic Ly$\alpha$ luminosity at the LAH's measured redshift, and then evaluated the selection function for the model from $z_{min}$ to $z_{max}$ in redshift steps equivalent to the MUSE spectral resolution (approximately $\Delta z = 0.001$ for the Ly$\alpha$ line center) at fixed intrinsic luminosity. 

We then evaluated 1/$V_{max}$ for each LAH in the L17 sample cut, integrating the selection function and volume element over the sample redshift range. 

\subsection{Fitting Intrinsic Parameter Distributions}

\begin{figure*}
    \centering
    \includegraphics[width=\textwidth]{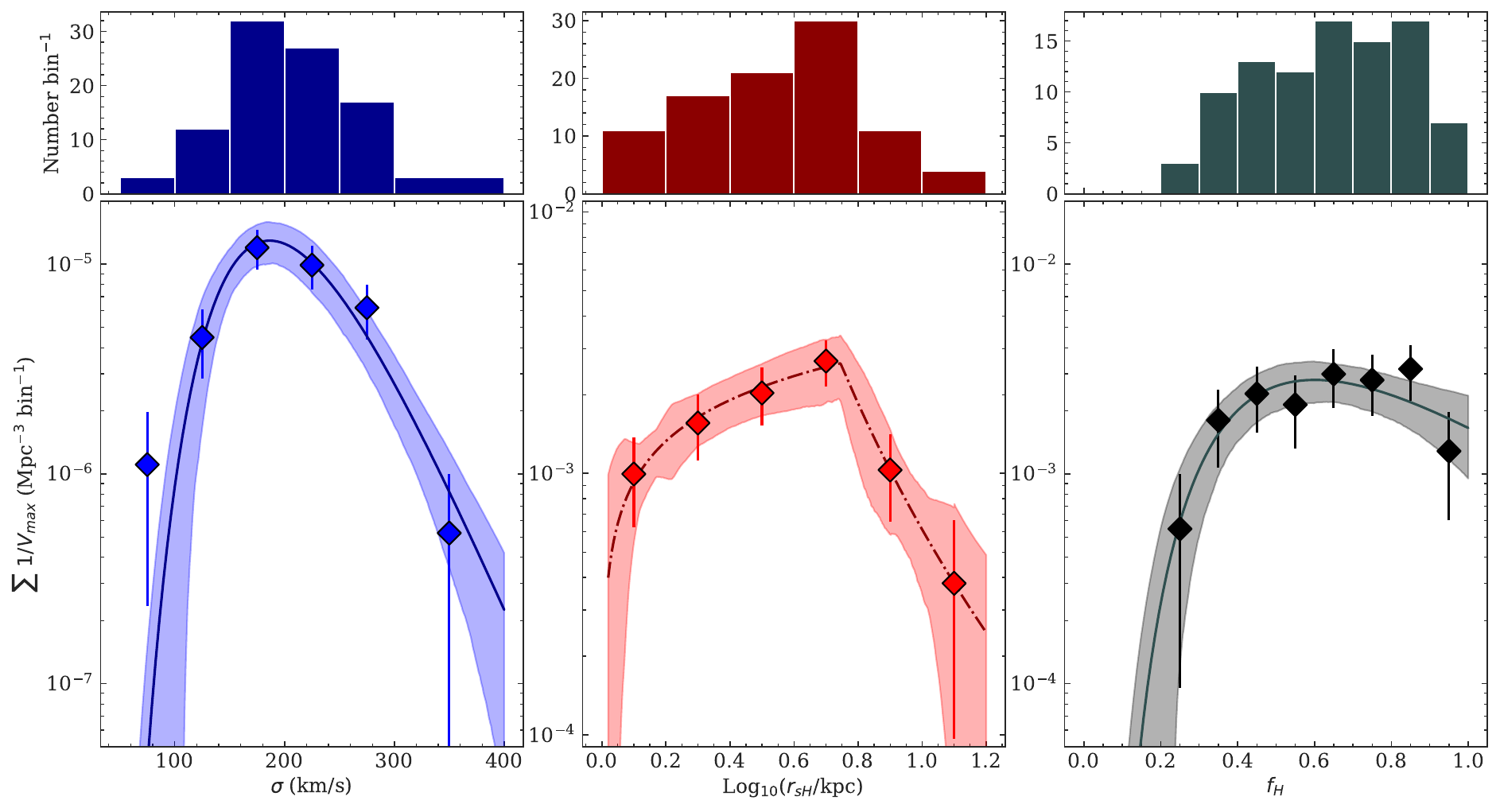}
    \caption{Differential parameter distributions for critical LAH physical characteristics. The panels show distributions for $\sigma$ (blue), $r_{sH}$ (red), and $f_H$ (black), with the top row giving a histogram of input sample LAHs, and the bottom panel showing the binned $1/V_{max}$ distributions. Individual bins are given by diamonds with error bars, with the best functional fit shown by the solid or dot-dash lines. Solid lines are lognormal fits, used for the $\sigma$ and $f_H$ distributions, while the $r_{sH}$ distribution was best fit by a broken power law (dot-dash line). The shaded region shows the inter-quartile range of possible fit parameters. The best-fit values are given in Table \ref{tab:fits}.}
    \label{fig:param_dist}
\end{figure*}

We evaluated Equation \ref{eq:vmax} for binned distributions of $\sigma$, $r_{sH}$, and $f_H$. The results are shown in Figure \ref{fig:param_dist}. Error bars for the 1/$V_{max}$ bins account for two major contributions to the uncertainty. First, we include an error term for the Poissonian counting statistics of the 1/$V_{max}$ sum \citep{johnston2011}. This is given by

\begin{equation} \label{eq:poisson}
\begin{centering}
    \sigma_{\Delta X_k} = \left[ \sum_{i=1}^N \left( \frac{1}{V_{max}} \right)^2 \right]^{1/2}
    \end{centering}
\end{equation}

The second major contribution to the error estimation results from errors in binning due to the measurement uncertainties of the halo parameters. We accounted for this with a simple bootstrap resampling, wherein we treat each parameter measurement in the sample as the center of an individual Gaussian distribution with a width equal to the measurement error. Then we resampled each measurement in the sample 1000 times, and performed the same binned 1/$V_{max}$ estimation for each of the 1000 new distributions. We took the standard deviation of the 1/$V_{max}$ sums in each bin to be the sampling error, and added this error in quadrature with the Poissonian statistics to represent the error in each bin. Typically, the sampling error was on the order of 10\% of the counting errors, so the error bars are dominated by the Poissonian term.

The $\sigma$ and $f_H$ distributions are both well-fit by a lognormal distribution:

\begin{equation} \label{eq:lognorm}
    \Phi(X,\mu,\nu,A) = \frac{A}{X \cdot \sqrt{2\pi\nu^2}} e^{-(\log_{10}X-\mu)^2/(2\nu^2)}
\end{equation}

where $X$ represents the mid-bin value for either $\sigma$ or $f_H$. The free parameters of the fit are the amplitude $A$, the mean $\mu$, and the standard deviation $\nu$. The best-fit values for these parameters are shown in Table \ref{tab:fits}. A lognormal function was chosen as the probability distribution that provided the best fit, though a Schechter function also provides a reasonable match.


\begin{table*}
\centering
\caption{Best-fit Lognormal and Smoothly Broken Power Law Distributions for Halo Parameters}
\begin{tabular}{lccccc}
\hline\hline
 $X_k$ & $\mu$ & $\nu$ & A \\
\hline
 $\sigma$ & $5.3\pm0.1$ & $0.27\pm0.04$ & $0.0017\pm3\times 10^{-4}$ \\
$f_H$ & $-0.26\pm0.22$ & $0.5\pm0.1$ & $0.0024\pm9\times 10^{-4}$ \\
\hline
\hline
$X_k$ & $x_b$ & A & $\alpha_1$ & $\alpha_2$ & $\Delta$ \\
\hline
$r_{sH}$ & $0.75\pm0.1$ & $0.0026\pm1 \times 10^{-3}$ & $-0.5\pm0.4$ & $5.0\pm3.9$ & $0.003\pm0.006$
\end{tabular}
\label{tab:fits}
\end{table*}

As with the grid construction, we analyze the 1/$V_{max}$ distribution as a function $r_{sH}$ in log space, allowing a finer sampling of the more-populated low-$r$ bins. This distribution we fit with a smoothly broken power law, defined as

\begin{equation} \label{eq:pl}
    \Phi(R_H,R_b,A,\alpha_1,\alpha_2,\Delta) = A \left( \frac{R_H}{R_b} \right)^{-\alpha_1} \left\{ \frac{1}{2} \left[ 1 + \left( \frac{R_H}{R_b} \right)^{\frac{1}{\Delta}} \right] \right\}^{(\alpha_1 - \alpha_2)\Delta}
\end{equation}

Here we define $R_H = \log_{10}(r_{sH})$, with free parameters $R_b$ (the ``break'' point between the two power laws), amplitude A, power law indices $\alpha_1$ and $\alpha_2$, and the smoothing parameter $\Delta$. Though we allowed these parameters to fully vary, we obtained the best fits only with a very small smoothing parameter of $\Delta=0.003$, just above the minimum allowable values of 0.001, resulting in a fit very similar to a simple broken power law.

For each fit, we obtained a confidence interval by repeating the fit over 1000 iterations, varying the values of the input 1/$V_{max}$ bins according to a Gaussian resampling with widths equal to the bin measurement errors. The shaded regions shown in Figure \ref{fig:param_dist} show the interquartile range of possible best-fit parameters from this process. We estimated uncertainty in the fit parameters as well by taking the inter-quartile range of their values from these 1000 perturbed fits. These are also shown in Table \ref{tab:fits}.

The general forms of the lognormal fits are well-constrained, with two caveats: one, the steepness of the dropoffs at low $\sigma$ or $f_H$ could be less constrained due to low-number statistics, and two, the dip as $f_H \rightarrow 1$ could be consistent with a flat distribution.

The confidence interval for the powerlaw fit shows more features. The bulge around $\log_{10}(r_{sH})\approx0.8$ and some of the other seemingly sharp features in the inverval are a result of variation in the free break parameter combined with the fits' tendencies toward very low smoothness. Compared with the lognormal fits, the edges of the distribution at very low and very high $\log_{10}(r_{sH})$ are less well constrained. However, the downturn at low $r_{sH}$ occurs below what is actually resolvable from the PSF, and so is not a meaningful prediction. At high $r_{sH}$, it is possible the distribution drops off much more steeply than predicted, but the overall configuration of rising then falling powerlaws stays the same.


\subsection{Implications for the Intrinsic LAH Population} \label{sec:imp}

Next we explore some of the implications of these findings for the intrinsic LAH population at $3 < z < 5$, beginning with the flux fraction and halo scale length. We discuss the line width and implications related to the spectral profile in \S\ref{sec:lsf}. First, we may confirm what was suggested above by the parameter tests and $S/N$ distributions: LAHs at $3<z<5$ with halo flux fractions $f_H < 0.3$ are intrinsically very rare for Ly$\alpha$ luminosities down to $10^{41.5}$ erg s$^{-1}$. As mentioned above in Sec. \ref{sec:sample}, this has been hinted at in previous observations that find low-$f_H$ galaxies only in small numbers and only at very high Ly$\alpha$ luminosities \citep[e.g.,][ in low-$z$ analogues]{runnholm2023}, an already-rare class of galaxy. But knowing the substantial detectability advantage such galaxies should have, we may now confirm their intrinsic rarity. This also supports previous indications that extended Ly$\alpha$ halos are essentially ubiquitous around LAEs at this redshift \citep{wisotzki2016, leclercq2017}.

Turning next to the intrinsic distribution of the scale lengths, we find that the most common halo scale length is expected to be around $\log_{10}(r_{sH})\approx0.7$, a scale length of about 5 kpc. This would mean the median halo is extended by factors of 2.5-5 more than the half-light radii measured for galaxies at these redshifts \citep{bouwens2004}. These scale lengths are not dramatically more common than LAHs with $r_{sH} < 4$ kpc, but it is notable that they are expected to be the most common scale length in the intrinsic distribution when the observed sample in Figure \ref{fig:hists} shows scale lengths of 1-3 kpc to be more frequent. From the parameter tests, we can expect LAHs with scale length of 2 kpc to be more detectable than those with $r_{sH} = 5$ by 0.1-0.2 dex in intrinsic flux, a distinction that shapes the difference between the observed and intrinsic distributions. 

The distribution drops off sharply at higher scale lengths, though the intrinsic distribution expects LAHs with $r_{sH} \approx 8$ kpc to be more common than in the observed distribution by about a factor of two. The slope of the dropoff is not well-constrained due to the very low number of $S/N$-sufficient halos, but such extreme extended objects are still expected to be quite rare. Nonetheless, this suggests the importance of the detection of low-surface-brightness galaxies in order to obtain a more complete observational sample. 

In our initial parameter tests, we operated under the assumption that each variable in the LAH profile could be studied in isolation, independent of the values of the other parameters. There is limited observational testing of this thus far, but L17 ($3<z<5$) and \citet[][low-$z$ analogues]{runnholm2023} find little indication that $f_H$ correlates meaningfully with the scale length. Relatedly, L17 and \citet{wisotzki2016} find a significant but relatively weak correlation between the halo and compact-component scale lengths (the Pearson correlation coefficient for the combined sample is $\rho=0.32$), albeit with high scatter. So although more extended halo components may be associated with more extended continuum-like components, it is still possible and even common to detect extended halos around more compact continuum-like sources. 


Observational tests relating these halo components to physical characteristics of the CGM or host galaxy are similarly limited, due to the small sizes of available samples of LAHs with fit parameters. Here we summarize some results in the literature concerning the halo scale length. With a sample of eight $3<z<6$ LAEs, \citet{song2020} find that the halo scale length is significantly driven by the underlying scale length of neutral hydrogen, determined as a fit parameter of the shell radiative transfer model. They find no significant correlation with other shell model parameters. The halo scale length may also correlate with characteristics of the stellar population, such as mass \citep{zhang2024}, age \citep{song2024}, and spatial extent of star-forming regions \citep{rasekh2022}, but the physical connection underlying these correlations must still be explored.

Finally, both simulations and clustering studies suggest that broad spatial extents of LAHs could be partially explained by contributions from faint LAE satellites \citep{byrohl2021, herrero2023b}. The L17 sample from which this intrinsic $r_{sH}$ distribution derives was selected to avoid satellites, so this should not be a contributing factor unless the surface brightness contribution from faint satellites below detection limits is significant. The possible impact of such faint satellites will be explored in subsequent work (Kozlova et al., in prep.).

\subsection{LSF Correction to the Line Width Distribution} \label{sec:lsf}

\begin{figure}
    \centering
    \includegraphics[width=0.5\textwidth]{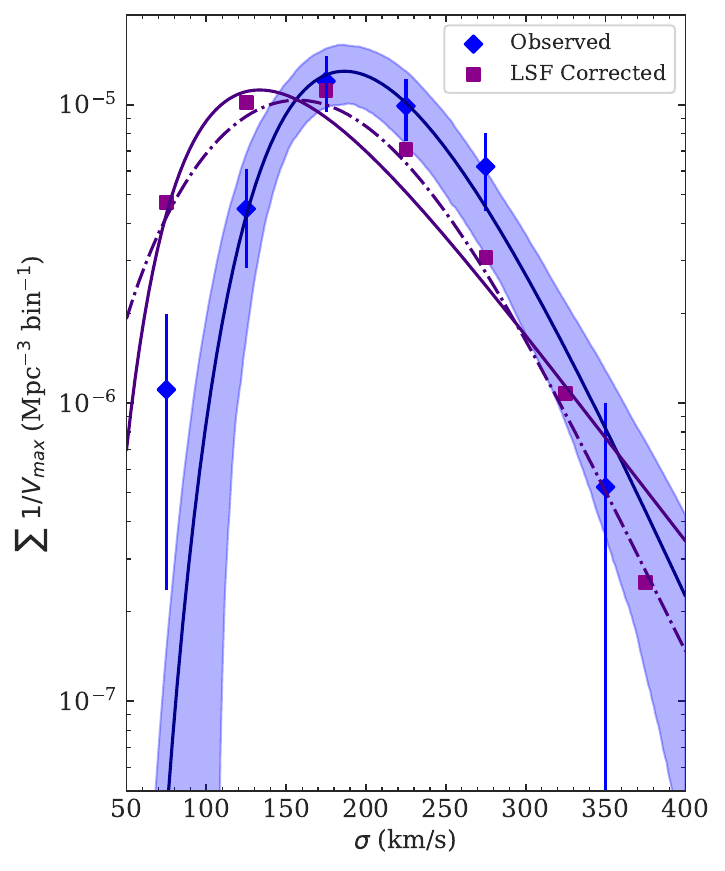}
    \caption{The intrinsic distribution of the line width $\sigma$ in both observed and LSF-corrected forms. Blue diamonds and the blue line and shaded region replicate the distribution from Fig. \ref{fig:param_dist}, which is the intrinsic distribution of the observed line widths. This represents a convolution of the physical spread of the line emission with the wavelength-dependent LSF. The purple squares show the intrinsic distribution de-convolved from the LSF, as described in Section \ref{sec:lsf}. While the observed $\sigma$ distribution is well-fit by a lognormal function, for the LSF-corrected distribution (shown with the purple solid line), this tends to produce a large tail at the high-$\sigma$ end that is unrealistic for a procedure that reduces the values of $\sigma$. A smoothly broken powerlaw function (dot-dash line), as used for the $r_{sH}$ distribution, models the tail end of the distribution more successfully. The LSF de-convolution reduces the $\sigma$ value of the peak of the distribution from 187 km/s to either 133 km/s (lognormal) or 157 km/s (powerlaw).}
    \label{fig:lsf_corr}
\end{figure}

\begin{table*}
\centering
\caption{Line Width Distribution Fit Parameters}
\begin{tabular}{ccccccc}
\hline\hline
Lognormal & Distribution & $\mu$ & $\nu$ & A & & \\
\hline
 & Observed & 5.3 & 0.267 & 0.00168 & & \\
 & Physical & $5.07\pm0.03$ & $0.42\pm0.02$ & $0.00175\pm1.4\times 10^{-4}$ & & \\
 \hline
 Powerlaw & Distribution & A & $\sigma_B$ & $\alpha_1$ & $\alpha_2$ & $\Delta$ \\
     &  Physical & $2.4 \times 10^{-6}\pm1.7 \times 10^{-6}$ & $287\pm35$ & $-2.1\pm0.13$ & $14\pm4$ & $0.31\pm0.03$ 
\end{tabular}
\label{tab:sig}
\end{table*}

Before commenting further on the implications of the intrinsic distribution of spectral line widths, we note that this quantity is not yet measuring a fully physical parameter of the galaxies. In observations, the physical size and shape properties of galaxies are convolved with the effects of observing conditions that alter the observed measurements, spatially through the point-spread function (PSF) and spectrally through the line-spread function (LSF). The spatial components such as $r_{sH}$ have already incorporated the PSF information through the Galfit procedure (see Sec. \ref{sec:modprof}), but spectral components such as $\sigma$ are still convolved with the LSF. 

This LSF was already incorporated into the LAH models used to generate the selection function grid for the UDF Mosaic, such that $\sigma = \sqrt{\sigma_0^2 + \sigma_{LSF}^2}$. This allowed us to model completeness fractions for observed properties of the L17 LAH sample, for which the $\sigma$ values fit in B23 give the observed line widths, including the LSF contribution.

For a specific sample of LAHs, such as the L17 subsample, by knowing the emission line wavelengths one could measure the LSF contribution and thereby determine the physical line spread ($\sigma_0$) for each galaxy. For the recovered intrinsic distribution of $\sigma$, however, this is not so straightforward. We could not directly measure the LSF contribution of a galaxy drawn from a predicted distribution, but with our knowledge of the redshift distribution of the observed sample, we could model it.

We took cumulative distributions of the redshifts of the L17 sample and the intrinsic distribution of $\sigma$ values shown in Figure \ref{fig:param_dist}, which we then normalized into cumulative density distributions. These we sampled one million times, drawing random redshifts and line widths for a mock sample of one million LAHs with statistically similar $z$ and $\sigma$ distributions.

For this mock sample, since each galaxy was assigned a specific redshift, we could calculate the LSF contribution to its line width and remove it from each mock LAH's $\sigma$ value, thereby obtaining $\sigma_0$ for that mock galaxy. With each mock galaxy deconvolved from its LSF, we took the numerical $\sigma_0$ distribution and rescaled it to the intrinsic distribution of observed line widths, thereby reconstructing an intrinsic distribution of physical line widths for $3<z<5$ LAHs.

This reconstructed distribution is shown in Figure \ref{fig:lsf_corr}, in comparison with the observed intrinsic distribution. Removing the LSF contribution obviously shifts the distribution toward lower $\sigma_0$ values, but we also found that, while the lognormal distribution well-described the observed intrinsic distribution, when applied to the physical $\sigma_0$ distribution, the fit tended to produce too large a tail at high-$\sigma$ values. As can be seen in the figure, the lognormal fit is unable to replicate the steeper dropoff at high $\sigma_0$, instead predicting slightly higher numbers of $\sigma_0 >300$ km/s lines in the deconvolved sample than in the uncorrected distribution, an unrealistic scenario.

We obtained a superior fit with the smoothly broken power law, previously applied to the $r_{sH}$ distribution. This function is better able to represent the expected steep drop off at high line widths, while still matching the more populated parts of the distribution. 

This choice of functional form does impact the interpretation of the physical line width distribution. Comparing the observed intrinsic fit to the two fits of the intrinsic physical distribution, we find that the peak line width in the LAH population reduces from 187 km/s in the observed intrinsic to 133 km/s (lognormal) or 157 km/s (powerlaw) in the physical reconstruction. The shift of 30 km/s from the observed to physical distributions is very similar to the results of Claeyssens et al. (in prep.), who performed a similar analysis of a sample of lensed LAEs and found an average correction of 20 km/s. 

Compared with the spatial parameters discussed above, the Ly$\alpha$ line width has had relatively more existing analysis in both observations and simulations. The intrinsic Ly$\alpha$ line width\footnote{In radiative transfer studies of Ly$\alpha$, the intrinsic line width typically refers to the width of the line prior to resonant scattering in the CGM, making this a distinct measure from the observed Ly$\alpha$ line width or even the LSF-deconvolved physical line width described above.} is a key component of the Ly$\alpha$ shell model \citep{verhamme2006, gronke2015}, though its value is often degenerate with other parameters of the model, such as the neutral hydrogen column density. \citet{yang2017} and \citet{hu2023} fit Ly$\alpha$ emission from low-redshift LAEs, including ``green pea'' galaxies, to the shell model, obtaining fit results for the shell model parameters, including the width of the Ly$\alpha$ line. Both are small samples, and both distributions peak at higher line widths (200-250 km/s) than the peaks predicted by the LSF-corrected intrinsic distributions. Subsequent studies of the shell model have noted that it predicts unrealistically broad intrinsic Ly$\alpha$ lines compared to observed Balmer emission \cite{orlitova2018}, indicating that the physically simplistic shell model is not a sufficient representation of the real CGM physics driving LAH spectral properties.

More recent attempts to model Ly$\alpha$ radiative transfer replace the expanding spherical shell model with a clumpy, multiphase medium \citep{li2021, li2022a}. \citet{li2022b} compare the two approaches, demonstrating both the existing parameter degeneracies in the shell model and showing that a clumpy model can better fit the broad wings and asymmetric profiles found in observed Ly$\alpha$ spectra without excessively broad intrinsic line widths. Using spatially resolved Keck/KCWI spectroscopy of 12 $z\sim2$ LAEs, \citet{erb2023} apply a clumpy, multiphase model to the observed Ly$\alpha$ profiles. They show that the best-fit clump velocity dispersions (one of the six critical parameters in the clumpy model) are found to be $\lesssim 150$ km/s, which is in line with both the LSF-corrected nebular line widths of their sample and comparable to the peak of our recovered physical line width distribution. In the multiphase clumpy model, intrinsic line widths (prior to radiative transfer) are small, and the velocity dispersion in the neutral gas clumps in the CGM are responsible for broadening the line width in the spectrum. This is a promising connection, and application of this clumpy model to larger LAE samples will further clarify the relationship between the intrinsic distribution of observed line widths and the kinematics of gas clouds in the CGM.

Apart from radiative transfer modeling, many studies have attempted to link Ly$\alpha$ spectral profile characteristics to other observable parameters of LAEs. \citet{yang2017} find that the width of the observed Ly$\alpha$ red peak (analogous to the spectral feature in our models) correlates with the Ly$\alpha$ escape fraction, and thus anticorrelated with the O32 ratio\footnote{Commonly defined as $\log_{10}\left( [\textsc{OIII}4959,5007]/[\textsc{OII}]3727,3729 \right)$}, a common diagnostic tracer of ionization in the interstellar medium. This anti-correlation is also observed weakly with a larger sample of 87 low-$z$ LAEs in \citet{hayes2023}, who also observe a weak correlation between O32 and the skewness parameter, but with peak line widths in agreement with the expectation from our distribution (see Figure 11 in that paper). This could imply a connection between the scattering line width of Ly$\alpha$ emission and the prevalence of ionization channels or ionized outflows in LAEs, although some recent Ly$\alpha$ escape models suggest the photons may be more prone to escape via scattering through higher-density gas than through ionized regions \citep{monter2024}. Ultimately, these are speculative connections in low-$z$ analogues; better comparisons via simulations or high-redshift observations are needed to disentangle the intrinsic line width's relationship to other characteristics of the halo and galaxy.

There do exist some observational comparisons at higher redshift. \citet{gonzalez2023} find very large ($\sim$500 km/s) line widths associated with strong, large-scale gas outflows in $z\sim4$ LAEs with QSO contributions, but these would be predicted to be an extreme rarity by our intrinsic distribution. L17 also inspected the relationship between the line width and the halo scale length. They measured no overall correlation, but did find that LAHs with $r_{sH} < 2$ kpc were found to only have narrow lines, while more extended LAHs had line widths across the whole parameter space. \citet{leclercq2020} measure a significant correlation between line width in the halo component and the halo flux fraction, and a weaker correlation between $f_H$ and the ratio of line widths in the halo and the compact component. While not a direct comparison, \citet{claeyssens2022} also find a significant correlation between the line width and the Ly$\alpha$ 50\% light radius. \citet{verhamme2018} note a correlation between Ly$\alpha$ line width and the line's velocity offset from systemic redshift among both low- and high-redshift LAEs, and \citet{muzahid2020} find for $z\sim3.3$ LAEs that this velocity offset also correlates with the host galaxy star formation rate (though this is not found in all studies, e.g. \citet{song2024}). This further hints at connections between the internal dynamics of the host galaxy, the escape pathways available to Ly$\alpha$ emission, the physical extension of the halo gas, and the observed properties of the line. We will explore these possible connections in subsequent work.



\section{Conclusions}

In this paper, we developed a 3D model of the spatial and spectral profiles of $3<z<5$ Lyman-$\alpha$ halos (LAHs) based on six key halo characteristics: the halo and compact exponential scale lengths ($r_{sH}$ and $r_{sC}$), the halo flux fraction ($f_H$), the compact component ellipticity ($q$), the spectral line width ($\sigma$), and the spectral line skewness parameter ($\gamma$). By inserting the model halos into miniature datacubes that mimic specific observations with VLT/MUSE, including the associated variance spectrum, we were able to test detection recovery of different LAH models with the Line Source Detection Cataloguing tool (LSDCat) as a function of the line central wavelength and the intrinsic Ly$\alpha$ line flux.

We used this procedure to test the impact on halo detectability of each of the six key parameters in isolation, finding that the line width $\sigma$, the halo scale length $r_{sH}$, and the halo flux fraction $f_H$ are the parameters expected to most influence the line detectability. 

We used a large sample of 145 Ly$\alpha$ emitters with measured halo spatial properties from the \citet{leclercq2017} analysis of deep MUSE observations in the UDF Mosaic, reaching $L_{Ly\alpha} > 41.5$ erg s$^{-1}$. Combining this with spectral properties from the MUSE Hubble Ultra Deep Field Data Release II \citep{bacon2023}, we had observational measures for the six LAH properties needed for the halo model. We designed a grid of models to span the observed parameter spaces, with finer resolution on the axes of the three most impactful parameters. With this spanning grid, we then developed a generalized LAH completeness model that marginalized over the distributions of the ellipticity, compact scale length, and skewness parameter. This allowed the construction of general selection functions for any LAH based on input line width, halo scale length, and halo flux fraction.

Taking a subset of the L17 halos for which the subsample was complete across the range of the three key parameters, we reconstructed the intrinsic distributions of these parameters using a version of the $1/V_{max}$ estimator that accounts for variable completeness as a function of intrinsic luminosity and redshift. We present the best-fit functional forms for the intrinsic distributions, finding that the $\sigma$ and $f_H$ distributions are well-represented by lognormal functions, while the $\log_{10}(r_{sH}/\text{kpc})$ distribution is best-fit by a smoothly-broken power law with a break at $\log_{10}(r_{sH}/\text{kpc}) = 0.75$. We confirm the intrinsic rarity of LAEs with low halo fractions $f_H < 0.3$ in this redshift-luminosity regime, and show that the most common halo scale lengths are toward the middle of the observed distribution ($r_{sH} \approx 5$ kpc), even though halos with smaller scale lengths are most common in the observations ($r_{sH} < 3$ kpc). This shows that LAHs tend to be more extended than observed distributions would indicate at first glance, and thus that analyses of LAH populations should carefully account for this less-detectable group.

We modelled an LSF deconvolution for the intrinsic $\sigma$ distribution, developing a distribution representing the physical line widths without the effects of observational line broadening. This intrinsic physical distribution was best fit by a smoothly broken power law, and reduced the peak line width from 187 km/s to 157 km/s. We compare these distributions to some basic predictions from shell models, as well as observations from L17 and of low-redshift Ly$\alpha$ analogues. There is some evidence of a correlation between the line width and ionization indicators in the interstellar gas, perhaps showing a connection between Ly$\alpha$ escape pathways and the physical extension of the emission line, but this study is still very limited and will have to be improved in subsequent work.

The methods we outline in this work will be applicable to other MUSE surveys, or indeed any contemporary or future IFU observations whose 3D profiles and noise properties can be determined in a similar way. It also provides a framework for modeling/predicting detection results for future Ly$\alpha$ surveys, such as what will be available with the upcoming BlueMUSE instrument, which may be able to detect much larger samples of LAHs in similar observing conditions to MUSE at lower redshift.

\begin{acknowledgements}

We thank the anonymous referee for their helpful suggestions. JP and LW acknowledge funding by the Deutsche Forschungsgemeinschaft, Grant Wi 1369/31-1. TU acknowledges funding from the ERC-AdG grant SPECMAP-CGM, GA 101020943. R.B. and L.W. acknowledge support from the ANR/DFG grant L-INTENSE (ANR-20-CE92-0015, DFG Wi 1369/31-1). Tran Thi Thai was funded by Vingroup JSC and supported by the Master, PhD Scholarship Programme of Vingroup Innovation Foundation (VINIF), Institute of Big Data, code VINIF.2023.TS.108. The authors would like to thank Constanza Mu\~{n}oz L\'{o}pez for helpful consultation on proper parentheses placement.
      
\end{acknowledgements}

%
%

\bibliographystyle{aa}
\bibliography{lae_refs}

\end{document}